\documentclass[conference,compsoc]{IEEEtran}

\usepackage[noend]{algpseudocode}
\usepackage{amsmath} 
\usepackage{amssymb}
\usepackage{amsthm}
\usepackage{array}
\usepackage{balance}
\usepackage{booktabs}       
\usepackage[nocompress]{cite}         
\usepackage{colortbl}
\usepackage{comment}
\usepackage{enumitem}
\usepackage{float}
\usepackage[symbol]{footmisc}
\usepackage{framed}
\usepackage[acronym,nomain]{glossaries}
\usepackage[pdftex]{graphicx} \graphicspath{ {./figures/} } 
\usepackage{hhline}
\usepackage{hyperref}
\usepackage{listings}
\usepackage{mathtools}
\usepackage{pgf-umlsd}      
\usepackage{msc} \mscset{msc keyword=}
\usepackage{multirow}
\usepackage{newtxmath}
\usepackage{orcidlink}
\usepackage{pdfpages} 
\usepackage{pgfplots} \pgfplotsset{compat=newest} 
\usepackage{pifont}
\usepackage{placeins}     
\usepackage{ragged2e}
\usepackage{soul}
\usepackage{siunitx}
\usepackage{stfloats}
\usepackage[caption=false,font=footnotesize,labelfont=sf,textfont=sf]{subfig}
\usepackage{svg}
\usepackage{tabularx}
\usepackage{tikz}
\usepackage{url}
\usepackage{xurl}
\usepackage{verbatim}
\usepackage{wasysym}
\usepackage{xcolor}
\usepackage{xfrac}

\makeatletter
\newcommand{\newlineauthors}{%
\end{@IEEEauthorhalign}\hfill\mbox{}\par
\mbox{}\hfill\begin{@IEEEauthorhalign}
}
\makeatother

\newtheorem{theorem}{Theorem}[section]

\newtheorem{proposition}[theorem]{Proposition}
\newtheorem{corollary}[theorem]{Corollary}

\theoremstyle{definition}

\theoremstyle{remark}

\makeatletter
\newcommand{\linebreakand}{%
  \end{@IEEEauthorhalign}
  \hfill\mbox{}\par
  \mbox{}\hfill\begin{@IEEEauthorhalign}
}
\makeatother

\IEEEoverridecommandlockouts

\begin{document}

\title{Don't Judge an Agent by Its Messages: \\Private-State Leakage under Passive Monitoring of Agent Communication}

\author{
  \IEEEauthorblockN{
    Daksh Pandey\IEEEauthorrefmark{2},
    Sayon Duttagupta\IEEEauthorrefmark{3},
    Xinhai Zou\IEEEauthorrefmark{3},
    Amit Singh Bhati\IEEEauthorrefmark{3}\IEEEauthorrefmark{4}
  }
  \IEEEauthorblockA{
    \IEEEauthorrefmark{2}IIT Roorkee\\
    \IEEEauthorrefmark{3}COSIC, KU Leuven\\
    \IEEEauthorrefmark{4}3MI Labs
  }
  \IEEEauthorblockA{
    daksh\_p@ph.iitr.ac.in,
    \{sayon.duttagupta, xinhai.zou\}@esat.kuleuven.be\\
    amitsingh.bhati@3milabs.tech
  }
}

\maketitle

\begin{abstract}

An AI agent can reproduce the distribution of its messages and of the receiver's task outcomes while disclosing private information that the task never required. We ask what a passive audit can certify when the auditor knows the authorised sender and the exact observable law but has no trusted pairing of messages with the private states behind them.
For a single round, compatible mutual-information leakage forms a sharp interval, and fixing the entire task-relevant message law, with every declared task utility, can still leave a full private bit undetermined; total leakage is determined if and only if, in every task state, the remaining private state or the authorised message is a deterministic function of that state. For a sender, such as an LLM, that selects among its own responses, we prove that in a balanced case its change to the message law within each task state equals $(1-2^{-k})$ times the honest messages' dependence on the attribute, which also sets, up to a constant, how many messages a monitor needs. With Qwen2.5-1.5B-Instruct on a synthetic routing task, selection among 16 responses lets a public decoder recover a private bit in all 8,192 held-out rounds at 98.60\% routing accuracy (honest sender 98.90\%). A monitor that knows the selection rule reaches AUC 0.645 on single messages, close to the ceiling of 0.671 that honest messages predict, and 0.972 over 16, while generic monitors stay near chance. A law-preserving sender built from the same samples recovers the bit in 97.7\% of rounds, and every monitor without the true bit stays near chance against it. A canonical task interface preserves every receiver decision and bounds the remaining information about a balanced attribute by twice the routing-error rate, even for a receiver that shares a key with the sender.
These results distinguish what passive auditing can certify from what restricting communication can enforce.

\end{abstract}

%
\IEEEpeerreviewmaketitle

\section{Introduction}\label{sec-introduction}

An inter-agent channel can reproduce its authorised message distribution exactly while carrying more information about the sender's private state. Consider a personal AI assistant running on a user's device. The assistant can access private emails, documents or health records, but may delegate a routine task to a cloud-based agent that has no legitimate need for unrelated information in the local files. A local agent that is compromised, for example through a tampered runtime or indirect prompt injection~\cite{greshake2023not}, or that colludes with the cloud agent, could condition its messages on that information and convey a sensitive fact, such as whether a private record contains a particular diagnosis, without stating it explicitly. The cloud agent would still complete its task, and the messages could keep the distribution expected at the communication boundary. This local-to-cloud trust boundary makes unwanted information flow and secret collusion a concrete concern for multi-agent systems~\cite{dewitt2025openchallenges,motwani2024secret}. In the terms of the confinement problem~\cite{lampson1973confinement}, the private fact leaves its compartment through a covert channel.

Monitoring is a natural response, but what can it establish? A passive auditor at the device's communication boundary can inspect every outgoing message, compare the distribution of these messages with that of an authorised sender, and observe the cloud agent's actions. It may even know how private records and task inputs are distributed. It is not told which record lies behind a given response, which would disclose that record, and network-level logs carry no trusted provenance linking a response to the private context used to generate it. This raises the question we study. \emph{What does agreement of the observed communication law certify about leakage of private state?} We grant the monitor deliberately favourable knowledge (the state distribution, the reference sender, the fixed receiver, the suspected attacks, and the exact observable law) but not the deployed state-conditioned sender or a trusted pairing of private state with message. Any ambiguity that remains is therefore structural, not a finite-sample limitation.

We make this concrete with the incident-routing task that we also use in our experiment. A sender reads an incident record, whose task-relevant part $R$ is one of $16$ routes, and writes a short message $Z$, from which a fixed receiver parses an action $A$. The record also contains administrative fields $N$ that routing does not need, one bit of which we call $B$. Sect.~\ref{subsec-concrete-construction} gives an exact pair of senders that produce the same joint distribution of route, message and receiver action and route every incident correctly, yet under one of them a public decoder recovers $B$ with certainty. Even a monitor that is told the true route of each incident cannot tell them apart.

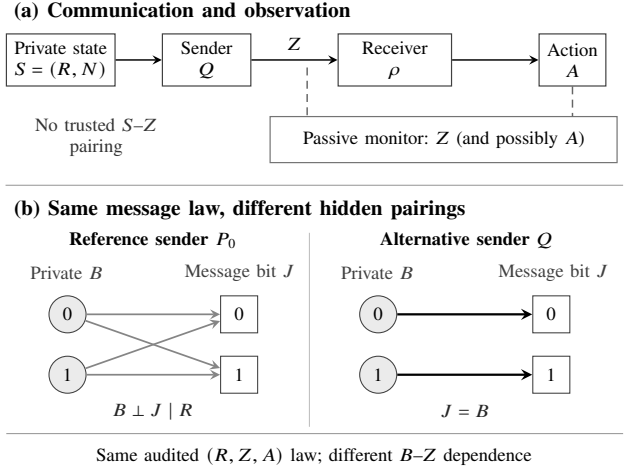
\begin{figure}[t]
    \centering
    \begin{tikzpicture}[x = 1cm, y=1cm,>=stealth,font=\scriptsize]
        \tikzset{agentbox/.style={draw=black!75,line width=.45pt,rectangle,fill=white,align=center,minimum height=.69cm,inner xsep=3pt,inner ysep=2pt}, channel/.style={->,draw=black,line width=.55pt}, auditlink/.style={draw=black!60,densely dashed,line width=.55pt}}
        \node[anchor=west,font=\footnotesize\bfseries] at (.07,.64) {(a) Communication and observation};
        \node[agentbox,minimum width=1.43cm] (s) at (.81,0) {Private state\\$S=(R,N)$};
        \node[agentbox,minimum width=1.17cm] (q) at (2.74,0) {Sender\\$Q$};
        \node[agentbox,minimum width=1.50cm] (rho) at (5.23,0) {Receiver\\$\rho$};
        \node[agentbox,minimum width=.88cm] (a) at (7.58,0) {Action\\$A$};
        \draw[channel] (s.east)--(q.west);
        \draw[channel] (q.east)--node[above] {$Z$}(rho.west);
        \draw[channel] (rho.east)--(a.west);
        \node[align=center,text=black!80] at (1.28,-1.02) {No trusted $S$--$Z$\\pairing};
        \draw[black!55,line width=.45pt] (3.58,-1.36) rectangle (8.17,-.75);
        \node[align=center] at (5.88,-1.06) {Passive monitor: $Z$ (and possibly $A$)};
        \draw[auditlink] (4.07,-.10)--(4.07,-.75);
        \draw[auditlink] (7.58,-.35)--(7.58,-.75);

        \draw[black!40,line width=.4pt] (.08,-1.66)--(8.25,-1.66);
        \node[anchor=west,font=\footnotesize\bfseries] at (.07,-1.98) {(b) Same message law, different hidden pairings};
        \node[font=\scriptsize\bfseries] at (2.03,-2.39) {Reference sender $P_0$};
        \node[font=\scriptsize\bfseries] at (6.18,-2.39) {Alternative sender $Q$};
        \node[text=black!75] at (.89,-2.82) {Private $B$};
        \node[text=black!75] at (3.18,-2.82) {Message bit $J$};
        \node[text=black!75] at (5.00,-2.82) {Private $B$};
        \node[text=black!75] at (7.33,-2.82) {Message bit $J$};
        \draw[black!25,line width=.4pt] (4.10,-2.29)--(4.10,-4.86);

        \tikzset{hidden/.style={draw=black!75,fill=black!8,circle,minimum size=.49cm,inner sep=0pt},observed/.style={draw=black!75,fill=white,rectangle,minimum size=.49cm,inner sep=0pt}}
        \node[hidden] (lb0) at (.90,-3.37) {$0$};
        \node[hidden] (lb1) at (.90,-4.17) {$1$};
        \node[observed] (lj0) at (3.17,-3.37) {$0$};
        \node[observed] (lj1) at (3.17,-4.17) {$1$};
        \node[hidden] (rb0) at (5.01,-3.37) {$0$};
        \node[hidden] (rb1) at (5.01,-4.17) {$1$};
        \node[observed] (rj0) at (7.30,-3.37) {$0$};
        \node[observed] (rj1) at (7.30,-4.17) {$1$};

        \draw[->,black!48,line width=.65pt] (lb0) -- (lj0);
        \draw[->,black!48,line width=.65pt] (lb0) -- (lj1);
        \draw[->,black!48,line width=.65pt] (lb1) -- (lj0);
        \draw[->,black!48,line width=.65pt] (lb1) -- (lj1);

        \draw[->,black,line width=.85pt] (rb0) -- (rj0);
        \draw[->,black,line width=.85pt] (rb1) -- (rj1);

        \node[text=black!85] at (2.03,-4.65) {$B\perp J\mid R$};
        \node[text=black!85] at (6.16,-4.65) {$J=B$};
        \draw[black!40,line width=.4pt] (.09,-4.95)--(8.25,-4.95);
        \node[align=center] at (4.16,-5.25) {Same audited $(R,Z,A)$ law; different $B$--$Z$ dependence};
    \end{tikzpicture}
    \vspace{-3pt}
    \caption{The missing private-state--message pairing. (a) A passive monitor observes messages and optionally receiver actions, but has no trusted pairing with the sender's private state. (b) In the exact witness of Sect.~\ref{subsec-concrete-construction}, both senders have the same task-conditioned message law $Z=(\widehat R,J)$: the reference draws $J$ independently of private bit $B$, whereas the alternative sets $J=B$. Edges indicate possible pairings, not their probabilities. The observable task law is unchanged despite different private-state leakage.}
    \label{fig-audit-boundary-roadmap}
        
\end{figure}

For a single round and an unrestricted sender, we characterise this ambiguity by treating the audit as a partial-identification problem. When only the overall message law $\beta$ is fixed, every coupling of the private-state distribution $\mu$ and $\beta$ is compatible with the audit, observing the actions of a fixed receiver adds nothing (Proposition~\ref{thm-observational-equivalence}), and the compatible leakage values form the sharp interval $[0,G(\mu,\beta)]$, where $G$ is the largest mutual information attained by such a coupling (Proposition~\ref{thm-leakage-interval}). The ambiguity survives a much stronger audit, in which the entire task-relevant communication law of $(R,Z)$ is fixed rather than an expected utility score. This preserves the full distribution of $(R,Z,A)$ and hence every declared utility based on $(R,A)$ (Proposition~\ref{prop-task-preserving-equivalence}), yet every task-conditioned coupling of a private attribute with the message remains realisable (Theorem~\ref{thm-task-preserving-coupling}). We derive the sharp leakage intervals and show that total leakage is determined if and only if, in every task state, the remaining private state or the authorised message is determined by that state (Corollary~\ref{cor-task-preserving-leakage}). We also give the best recovery of a private attribute by a public decoder, an exact condition for perfect recovery, and the best recovery when the message law may change by a bounded amount (Corollaries~\ref{cor-optimal-public-recovery} to~\ref{cor-distortion-recovery}). Fig.~\ref{fig-audit-boundary-roadmap} illustrates the audit boundary and two such senders.

The coupling viewpoint connects our analysis to information-theoretic steganography, where perfectly secure encodings correspond to couplings and efficient ones to minimum-entropy couplings~\cite{dewitt2023perfect}. We use the same tools from the auditor's side. The hidden content is then the sender's own private state, on which honest messages may already depend, and the question is which dependencies on it an audit can exclude, including under task preservation, and when it excludes them all. That an output law alone does not bound dependence on secrets is the probabilistic form of the refinement paradox of information flow~\cite{jacob1989derivation,morgan2006shadow}; our results quantify it for passive audits.

The formal sender model permits arbitrary state-conditioned message kernels. To show one realisable mechanism, we analyse an LLM sender that draws $k$ ordinary reference responses and emits the first whose public decoding agrees with a private bit, a bounded, keyless variant of the rejection-sampling encoder of provably secure steganography~\cite{hopper2002provably}. Its recovery law is exact, and in a balanced case its detectability is proportional to the honest sender's own dependence on the attribute (Theorem~\ref{thm-selection-detectability}), so a more discreet honest sender makes selection harder to detect. With Qwen2.5-1.5B-Instruct~\cite{yang2024qwen25} on a synthetic incident-routing task, selection among $k=16$ candidates recovers the private bit in all $8{,}192$ held-out rounds, at $98.60\%$ routing accuracy against $98.90\%$ for the honest sender (paired $95\%$ interval $[-0.65,0.04]$ points). Monitors based on message structure, character features and surprisal stay near chance (AUC $0.49$ to $0.50$), while a monitor that knows the selection rule reaches AUC $0.645$ on single messages, close to the ceiling of $0.671$ that Theorem~\ref{thm-selection-detectability} predicts from honest messages alone, and $0.972$ over $16$. The law-preserving sender of Proposition~\ref{prop-resampling-sender}, run on the same samples, recovers the bit in $97.72\%$ of rounds (predicted $98.00\%$), and that monitor gives only AUC $0.540$ against it. Mapping messages to a canonical form of the parsed route leaves every receiver decision unchanged and bounds what any decoder can learn about the bit in terms of the routing-error rate, also for a receiver that shares a key with the sender. The law-preservation results are proved and realised by the construction of Sect.~\ref{subsec-concrete-construction}, and the experiment shows their practical consequence.

\noindent\textbf{Contributions.}
\begin{itemize}
    \item \textbf{Sharp limits of passive auditing (Sects.~\ref{sec-leakage-identification} and~\ref{sec-utility-preserving}).} For a single round and an unrestricted sender, the leakage values compatible with an audit form sharp intervals, both for a fixed message law (Proposition~\ref{thm-leakage-interval}, Corollary~\ref{cor-other-measures}) and for a fixed task-relevant law that preserves every declared utility (Theorem~\ref{thm-task-preserving-coupling}, Corollary~\ref{cor-task-preserving-leakage}). Matching the task-relevant law determines total leakage if and only if, in every task state, the remaining private state or the authorised message is determined by that state. An exact witness preserves the full task-relevant observable law while leaking one more private bit (Table~\ref{tab-exact-witness}).
    \item \textbf{Optimal recovery (Sects.~\ref{subsec-leakage-consequences} and~\ref{subsec-selection-detectability}).} We give the best recovery of a private attribute by a public decoder and an exact condition for perfect recovery (Corollaries~\ref{cor-optimal-public-recovery} and~\ref{cor-perfect-public-recovery}), an idealised law-preserving sender that attains the optimum with two reference draws per message on average (Proposition~\ref{prop-resampling-sender}), and the best recovery when each task-conditioned message law may change by a bounded amount (Corollary~\ref{cor-distortion-recovery}).
    \item \textbf{Detectability of selection (Sect.~\ref{sec-selection-channels}).} For a sender, such as an LLM, that selects among its own responses, we give the exact kernel and recovery law and prove that, in a balanced case, its change to each task-conditioned message law is $(1-2^{-k})$ times the honest attribute--message dependence in total variation, and $(1-2^{-k})^2$ times it in $\chi^2$ divergence, which determines, up to a constant factor, how many messages a monitor needs (Theorem~\ref{thm-selection-detectability}). A more discreet honest sender thus makes selection harder to detect, while a sender that shares fresh key bits with the receiver can leave even the state--message law unchanged.
    \item \textbf{Measurements with an open-weight LLM (Sect.~\ref{sec-empirical-evaluation}).} Selection makes a private bit recoverable in all $8{,}192$ test rounds at $98.60\%$ routing accuracy against $98.90\%$ for the honest sender, and recovery matches the exact selection law within $0.55$ points at every budget. Generic monitors stay near chance, and a selection-aware monitor reaches AUC $0.645$ on single messages, close to the ceiling of $0.671$ that Theorem~\ref{thm-selection-detectability} predicts from honest messages, and $0.972$ over $16$. A law-preserving sender built from the same samples attains the predicted recovery while that monitor stays near chance.
    \item \textbf{Interface enforcement (Sects.~\ref{subsec-canonical-interfaces} and~\ref{subsec-empirical-canonicalization}).} A canonical task interface preserves every receiver decision by construction and, for an attribute balanced within each task state, keeps residual information at most twice the routing-error rate for every sender, including keyed ones (Proposition~\ref{prop-canonical-balanced}); on our test distribution this is $0.028$ bits.
\end{itemize}

Full proofs are in Appendix~\ref{app-proofs}, and experimental details and reproducibility information are in Appendix~\ref{app-experiments}.

\noindent\textbf{Artefacts.} An artefact bundle is provided at \url{https://anonymous.4open.science/r/passive-monitoring-EC82/}. It includes the code that reproduces the experiment in Sect.~\ref{sec-empirical-evaluation}, with the pinned model, seeds and sample sizes.

\section{Model}\label{sec-model}

We model one round of communication between a sender that observes a private state and a fixed receiver. The central distinction is between the sender's dependence on that private state and the distribution of messages that is visible to a passive monitor.

\subsection{Agents, Private State, and the Reference Sender}\label{subsec-system-model}

Let $\mathcal{S}$, $\mathcal{Z}$, and $\mathcal{A}$ be finite sets of private states, messages, and receiver actions. We condition throughout on a fixed public context and suppress it from the notation. The environment draws $S\sim\mu$, where zero-probability states are omitted from $\mathcal{S}$. The sender observes the realised state $S=s$ and emits a message $Z$ according to a stochastic kernel $Q(z\mid s)$, and a fixed receiver then produces an action according to $\rho(a\mid z)$. The resulting joint distribution is
\begin{equation}
    p_Q(s,z,a) = \mu(s)Q(z\mid s)\rho(a\mid z).
    \label{eq-joint-law}
\end{equation}

For an LLM-based sender, the kernel $Q$ includes any prompting, sampling, decoding, or output-selection procedure that affects the message sent; our analysis concerns this emitted law, not a particular implementation.

We compare the deployed sender with a fixed reference sender $P_0(z\mid s)$, whose induced message distribution is
\begin{equation}
    \beta(z) := \sum_{s\in\mathcal{S}}\mu(s)P_0(z\mid s).
    \label{eq-reference-message-law}
\end{equation}
We often refer to $P_0$ as the ``honest'' sender, but only in the sense of a prescribed baseline. The term does not imply that the reference communication is private or harmless: ordinary task-directed messages may already reveal information about the private state.

Unless stated otherwise, $Q$ ranges over all stochastic kernels from $\mathcal{S}$ to $\mathcal{Z}$. Particular model architectures or trusted output mechanisms may permit a smaller class; our theoretical results first ask what follows from the observation boundary alone.

\subsection{What the Monitor Sees and Knows}\label{subsec-auditor-model}

For a deployed sender $Q$, write
\begin{equation}
    \bar Q(z) := \sum_{s\in\mathcal{S}}\mu(s)Q(z\mid s)
    \label{eq-aggregate-msg-law}
\end{equation}
for the distribution of emitted messages. The baseline monitor observes $Z$ and may also observe the receiver action $A$. Since the receiver is fixed, their observable joint law is
\begin{equation}
    p_Q(z,a) = \bar Q(z)\rho(a\mid z).
    \label{eq-observable-joint-law}
\end{equation}
Thus, once the message law $\bar Q$ is fixed, observing the receiver action does not reveal anything further about which sender kernel produced it.

The monitor is passive, in the sense of the warden in Simmons's prisoners' problem~\cite{simmons1984prisoners}: it observes communication crossing the channel without altering it. We otherwise give it substantial knowledge. It knows the state distribution $\mu$, the reference sender $P_0$, and the receiver $\rho$; it may also know which attacks are suspected, though not which sender is deployed. For the negative results we also give it the exact distribution at its observation boundary, so that none of them can be attributed to sampling error.

What it does not receive is the realised private state associated with each deployed message. In particular, it has no trusted record pairing $S$ with $Z$, and no trusted interface through which it can query the deployed state-conditioned kernel $Q(\cdot\mid s)$. This distinction is the source of the problem studied below. Knowing how the reference sender would behave on a hypothetical state does not reveal which state produced an observed deployed message.

A monitor with trusted access to the realised private state has a strictly stronger observation boundary (Sect.~\ref{subsec-stronger-audit}). Our question is what a passive communication monitor of the kind described above can certify. Because it knows the exact laws and the suspected attacks, whatever it cannot certify is due to the missing state--message pairing.

\subsection{Leakage and the Compatible Set}\label{subsec-leakage-compatible-set}

We measure private-state information in the emitted message by the mutual information $I_Q(S;Z)$~\cite{cover2006elements}, where the subscript records that the joint law of $(S,Z)$ depends on the deployed sender $Q$, and compare the deployed sender with the reference through the leakage increment $\Delta(Q):=I_Q(S;Z)-I_0(S;Z)$, where the subscript $0$ denotes quantities computed under $P_0$.

The difficulty is that $I_Q(S;Z)$ depends on the joint state--message law, whereas the baseline monitor observes only its message marginal. To make this precise, write $\mathcal{P}(\mathcal{S}\times\mathcal{Z})$ for the set of probability distributions on $\mathcal{S}\times\mathcal{Z}$, and let
\begin{equation}
    \Gamma(\mu,\beta) := \left\{ \gamma\in\mathcal{P}(\mathcal{S}\times\mathcal{Z}) : \gamma_S=\mu,\; \gamma_Z=\beta \right\}
    \label{eq-couplings-set}
\end{equation}
be the set of couplings of the state distribution $\mu$ and the reference message distribution $\beta$. Define
\begin{equation}
    G(\mu,\beta) := \max_{\gamma\in\Gamma(\mu,\beta)} I_\gamma(S;Z).
    \label{eq-max-compatible-information}
\end{equation}
Equivalently,
\begin{equation}
    G(\mu,\beta) = H(\mu)+H(\beta) - \min_{\gamma\in\Gamma(\mu,\beta)}H(\gamma),
    \label{eq-mec-expression}
\end{equation}
so $G$ is attained by a minimum-entropy coupling of the two marginals, which is NP-hard to compute in general~\cite{kovacevic2015entropy} but approximable to within one bit in polynomial time~\cite{cicalese2019minimum}.

Finally, we call
\begin{equation}
    \mathcal{I}(\beta) := \left\{ I_Q(S;Z): \bar Q=\beta \right\}
    \label{eq-compatible-leakage-set}
\end{equation}
the \emph{compatible leakage set}: the set of private-state leakage values that remain possible after fixing the communication law observed by the monitor.

Sect.~\ref{sec-leakage-identification} characterises this set. We then strengthen the observation boundary by fixing how messages are distributed for the task-relevant part of the private state.

\section{Leakage Under a Fixed Message Law}\label{sec-leakage-identification}

We first consider the baseline monitor of Sect.~\ref{sec-model}. Suppose the deployed sender produces exactly the same message distribution as the reference sender,
\begin{equation}
    \bar Q = \beta.
    \label{eq-equal-message-law}
\end{equation}
This is much weaker than requiring $Q(\cdot\mid s)=P_0(\cdot\mid s)$ for every private state. Different state-conditioned sender kernels may have the same message marginal because marginalising over $S$ removes the information about which states produced which messages.

We now ask two questions. First, can the monitor distinguish such senders from the reference using the variables available at its audit boundary? Second, how much can their private-state leakage still differ?

\subsection{Observational Equivalence}\label{subsec-observational-equivalence}
Equal message laws already imply equal distributions over everything observed by the baseline monitor.

\begin{proposition}[Equal message laws give the same audit distribution]
    Let $Q$ satisfy $\bar Q=\beta$. Then $Q$ and the reference sender $P_0$ induce the same distribution over $(Z,A)$. Consequently, every randomised test or public post-processing of $(Z,A)$ has the same distribution under the two senders.
    \label{thm-observational-equivalence}
\end{proposition}

\begin{proof}
    By Eq.~\eqref{eq-observable-joint-law}, $p_Q(z,a)=\bar Q(z)\rho(a\mid z)$, which equals $\beta(z)\rho(a\mid z)=p_0(z,a)$ if $\bar Q=\beta$.
\end{proof}

Thus no test on the monitored variables distinguishes the two senders better than chance, and the law of any public decoder $d(Z)$ is fixed too, but not how the decoded value relates to the private state.

\subsection{Compatible Leakage Range}\label{subsec-compatible-leakage-range}

The preceding proposition concerns what the monitor can observe. It does not imply that the hidden state--message dependence is the same.

Recall from Sect.~\ref{sec-model} that fixing $\mu$ and $\beta$ leaves the joint law of $(S,Z)$ free to range over the coupling set $\Gamma(\mu,\beta)$. This gives the complete compatible leakage range.

\begin{proposition}[Leakage range under a fixed message law]
    Over the full sender class,
    \begin{equation}
        \mathcal{I}(\beta) = [0,G(\mu,\beta)].
        \label{eq-compatible-leakage-range}
    \end{equation}
    In particular, the message law determines the leakage if and only if $\mu$ or $\beta$ is a point mass.
    \label{thm-leakage-interval}
\end{proposition}

Every sender with $\bar Q=\beta$ induces the coupling $\gamma_Q(s,z)=\mu(s)Q(z\mid s)$ in $\Gamma(\mu,\beta)$, and, since $\mu(s)>0$, every coupling $\gamma$ defines such a sender $Q_\gamma(z\mid s)=\gamma(s,z)/\mu(s)$. The independent coupling gives zero mutual information, a minimum-entropy coupling attains $G(\mu,\beta)$, and interpolating between them attains every value in between (Appendix~\ref{app-proof-fixed-law}).

Because the reference sender itself has message law $\beta$, the corresponding range of changes relative to the reference is
\begin{corollary}[Leakage change relative to the reference sender]
    \begin{equation}
        \left\{ \Delta(Q):\bar Q=\beta \right\} = \left[ -I_0(S;Z), \; G(\mu,\beta)-I_0(S;Z) \right].
        \label{eq-compatible-increment-range}
    \end{equation}
\end{corollary}

The same holds for the vulnerability-based measures of quantitative information flow~\cite{alvim2020science}.
\begin{corollary}[Other leakage measures]\label{cor-other-measures}
For every gain function $g$ with positive prior vulnerability, the compatible multiplicative $g$-leakage values over senders with $\bar Q=\beta$ form an interval from $0$ to a maximum attained at a vertex of $\Gamma(\mu,\beta)$. For Bayes vulnerability this maximum is $\log_2\bigl(\max_d[1-\mathrm{TV}(\mu,q^d)]/\max_s\mu(s)\bigr)$, where $d$ ranges over maps $\mathcal{Z}\to\mathcal{S}$ and $q^d$ is the law of $d(Z)$ under $\beta$.
\end{corollary}

Evaluating this maximum is NP-hard in general, since for $\mu$ uniform on two states it equals $1$ if and only if $\beta$ gives some set of messages mass $1/2$, which is the partition problem.

The important distinction is therefore between fixing what the monitor sees and fixing how those observations depend on private state. Equal message laws fix the former exactly and leave the latter undetermined unless $\mu$ or $\beta$ is a point mass.

\subsection{Composition Over Rounds}\label{subsec-composition-rounds}

The same reasoning applies to repeated communication, but the object that must be preserved is the distribution of the \emph{entire transcript}, not merely the message distribution in each round separately.

For $T$ rounds, write $S^{T}=(S_1,\ldots,S_T)$ and $Z^{T}=(Z_1,\ldots,Z_T)$, let $\mu^{(T)}$ be the law of $S^T$, and let $\beta^{(T)}(z^{T}):=\Pr_0[Z^{T}=z^{T}]$ be the reference law of the transcript. Treating $S^{T}$ and $Z^{T}$ as single variables, Propositions~\ref{thm-observational-equivalence} and~\ref{thm-leakage-interval} apply to the unrestricted \emph{block-sender} class, in which the sender may observe the entire sequence $S^T$ before generating $Z^T$. In particular, if a deployed system satisfies
\begin{equation}
    \Pr_Q[Z^{T}=z^{T}] = \beta^{(T)}(z^{T}) \qquad\forall z^{T},
    \label{eq-transcript-law-equality}
\end{equation}
then its monitored message transcript has the same distribution as under the reference sender. Over the unrestricted block-sender class, the compatible leakage values are exactly
\begin{equation}
    I_Q(S^{T};Z^{T}) \in \left[ 0, G\!\left(\mu^{(T)},\beta^{(T)}\right) \right].
    \label{eq-multiround-leakage-range}
\end{equation}
For a \emph{causal} sender, whose message $Z_t$ may depend only on $S^t$ and $Z^{t-1}$, Eq.~\eqref{eq-multiround-leakage-range} remains an outer bound, but not necessarily an attainable range: not every coupling of $S^T$ and $Z^T$ can be implemented without access to future states. $G$ is also superadditive over rounds. For a fair bit and $\beta=(2/3,1/3)$, one round admits $0.459$ bits, but two independent rounds admit $1.394$ bits for a block sender and $1.099$ for a causal one. Causal senders correspond to the couplings in which each $Z^t$ is independent of $(S_{t+1},\ldots,S_T)$ given $S^t$; with $\mu^{(T)}$ fixed these constraints are linear, so their compatible leakage values also form an interval, with the maximum at a vertex.

Matching each round's marginal is not sufficient either. Two fair message bits that are independent under the reference but equal under an alternative sender have the same round marginals, yet a monitor that sees both rounds tells them apart, since $Z_1=Z_2$ with probability $1/2$ under one and $1$ under the other.

If rounds are genuinely independent under both systems, then equality of the single-round law does compose by factorisation. In a stateful or adaptive system, or for a private attribute that persists across rounds, such independence cannot be assumed, and the relevant audit object is the full transcript law.

\section{Leakage Under a Fixed Task-Relevant Law}\label{sec-utility-preserving}

Sect.~\ref{sec-leakage-identification} fixed the overall message law, but some sender kernels satisfying that condition may no longer communicate what the receiver needs for its task. We now impose a stronger requirement: preserve how messages are distributed for each task-relevant state. We show that this preserves the declared task behaviour while still allowing the message to carry different information about the remaining private state.

\subsection{Task-Relevant State and Utility}\label{subsec-task-residual-state}

Let the private state be $S=(R,N)$, where $R$ contains the information relevant to the declared task and $N$ is the remaining private state. We do not assume that they are independent. Write $\pi(r)=\Pr[R=r]$ and $\alpha_r(n)=\Pr[N=n\mid R=r]$ for $\pi(r)>0$. We consider a bounded task utility $u(R,A)$, with
\begin{equation}
    U(Q):=\mathbb{E}_Q[u(R,A)].
    \label{eq-expected-utility}
\end{equation}
Thus, the declared task objective depends on the private state through $R$ and on the receiver's action $A$.

Under the reference sender, define
\begin{equation}
    \beta_r(z):=\Pr_0[Z=z\mid R=r],\qquad
    \lambda_0(r,z):=\pi(r)\beta_r(z).
    \label{eq-honest-task-conditioned-joint}
\end{equation}
Here $\lambda_0$ is the reference joint law of $(R,Z)$. Unlike the overall message law $\beta$, it records how communication varies with the task-relevant part of the private state.

\subsection{Task-Preserving Class}\label{subsec-task-preserving-class}

We restrict attention to senders that reproduce this joint law:
\begin{equation}
    \mathcal{F}_R:=\left\{Q:P_Q^{(R,Z)}=\lambda_0\right\}.
    \label{eq-task-preserving-class}
\end{equation}
Equivalently, $\Pr_Q[Z=z\mid R=r]=\beta_r(z)$ for every $r$ of positive probability. This is stronger than matching the overall message distribution.

\begin{proposition}[Preservation of task-relevant behaviour]
    Every $Q\in\mathcal{F}_R$ preserves the reference joint distribution of $(R,Z,A)$. In particular, $\bar Q=\beta$ and $U(Q)=U(P_0)$.
    \label{prop-task-preserving-equivalence}
\end{proposition}

Indeed, for every $Q\in\mathcal{F}_R$,
\begin{equation}
    \Pr_Q[R=r,Z=z,A=a]
    =\lambda_0(r,z)\rho(a\mid z),
    \label{eq-task-preserving-rza-law}
\end{equation}
which is also the reference law. The declared utility depends only on $(R,A)$, so it is unchanged. This condition is sufficient, but not necessary, to preserve expected utility: different communication laws can give the same value of the scalar objective $U(Q)$.

\subsection{Attribute--Message Couplings Under a Fixed Task Law}\label{subsec-complete-characterize}

Fix any finite-valued private attribute $B=g(R,N)$ and let $p_r(b)=\Pr[B=b\mid R=r]$. Within each task-relevant state $r$, the attribute distribution $p_r$ and the message distribution $\beta_r$ are fixed. The following result shows that these are the only restrictions on their joint dependence.

\begin{theorem}[Attribute--message couplings under a fixed task law]
    For every $r$ with $\pi(r)>0$,
    \begin{equation}
        \left\{P_Q^{(B,Z)\mid R=r}:Q\in\mathcal{F}_R\right\}
        =\Gamma(p_r,\beta_r).
        \label{eq-task-preserving-coupling-complete}
    \end{equation}
    Any choice of one coupling from $\Gamma(p_r,\beta_r)$ for each $r$ can be realised simultaneously by a sender in $\mathcal{F}_R$.
    \label{thm-task-preserving-coupling}
\end{theorem}

To see why, choose couplings $\eta_r\in\Gamma(p_r,\beta_r)$ and, on the positive state support, define
\begin{equation}
    Q_\eta(z\mid r,n)
    :=\frac{\eta_r(g(r,n),z)}{p_r(g(r,n))}.
    \label{eq-coupling-induced-encoder}
\end{equation}
This is a valid sender kernel. Summing over $N$ gives $\Pr_{Q_\eta}[Z=z\mid R=r]=\sum_b\eta_r(b,z)=\beta_r(z)$, while its conditional joint law of $(B,Z)$ is exactly $\eta_r$. The full verification is in Appendix~\ref{app-proof-task-coupling}.

Thus, preserving the task-relevant communication law does not fix how messages depend on a selected private attribute. Taking $B=N$ gives the same characterisation for the entire remaining private state. If $R$ is constant, this reduces to the fixed-message-law setting of Sect.~\ref{sec-leakage-identification}.

\subsection{Leakage and Recovery}\label{subsec-leakage-consequences}

The coupling characterisation gives exact leakage ranges. Define
\begin{equation}
    I_R:=I_0(R;Z),\qquad
    C_R:=\sum_r\pi(r)G(\alpha_r,\beta_r),
    \label{eq-residual-leakage-capacity}
\end{equation}
and, for the selected attribute,
\begin{equation}
    C_B:=\sum_r\pi(r)G(p_r,\beta_r).
    \label{eq-targeted-leakage-capacity}
\end{equation}

\begin{corollary}[Leakage under task-law preservation]
    Over $\mathcal{F}_R$, the possible total and attribute-specific leakage values are exactly
    \begin{align}
        \{I_Q(S;Z):Q\in\mathcal{F}_R\}
        &=[I_R,I_R+C_R],
        \label{eq-task-preserving-leakage-interval}\\
        \{I_Q(B;Z\mid R):Q\in\mathcal{F}_R\}
        &=[0,C_B].
        \label{eq-targeted-leakage-interval}
    \end{align}
    Moreover, $C_R=0$ if and only if, for every $r$ with $\pi(r)>0$, $\alpha_r$ or $\beta_r$ is a point mass, and $C_B=0$ if and only if, for every such $r$, $p_r$ or $\beta_r$ is a point mass.
    \label{cor-task-preserving-leakage}
\end{corollary}

The chain rule gives $I_Q(S;Z)=I_R+I_Q(N;Z\mid R)$. Within each $r$, independence attains the lower endpoint and a maximum-information coupling attains the upper endpoint. Continuity gives every intermediate value. The same argument applies to $B$, and $C_B\leq C_R$ because $B$ is a function of $(R,N)$. Appendix~\ref{app-proof-task-leakage} proves attainability of every intermediate value in both intervals and the point-mass conditions.

The reference may already leak information about $N$. Writing $K_0:=I_0(N;Z\mid R)$, the possible increments are
\begin{equation}
    \{\Delta(Q):Q\in\mathcal{F}_R\}
    =[-K_0,C_R-K_0].
    \label{eq-task-preserving-relative-leakage}
\end{equation}
Hence the task-relevant law determines total leakage if and only if, in every route, $N$ or the reference message is determined by $R$, as for an error-free canonical interface (Sect.~\ref{subsec-canonical-interfaces}), and a strictly positive increase over the reference is possible when $C_R>K_0$.

Leakage is not the same as successful recovery of an individual attribute. Consider a fixed, public decoder $d:\mathcal{Z}\to\mathcal{B}$, and let
\begin{equation}
    q_r^d(b):=\sum_{z:d(z)=b}\beta_r(z).
    \label{eq-decoder-dist}
\end{equation}
For distributions $p,q$, write $\operatorname{TV}(p,q)=\tfrac12\sum_b|p(b)-q(b)|$.

\begin{corollary}[Recovery with a public decoder]
    For any fixed $d$,
    \begin{equation}
        \begin{aligned}
        \max_{Q\in\mathcal{F}_R}\Pr_Q[d(Z)=B]
        &=\sum_r\pi(r)\bigl[1-\operatorname{TV}(p_r,q_r^d)\bigr].
        \end{aligned}
        \label{eq-optimal-public-recovery}
    \end{equation}
    \label{cor-optimal-public-recovery}
\end{corollary}

Within each $r$, a maximal coupling of $B$ and $d(Z)$ attains the stated agreement probability~\cite[Prop.~4.7]{levin2017markov}; Theorem~\ref{thm-task-preserving-coupling} then realises it through a sender with the required message law (Appendix~\ref{app-proof-public-recovery}).

\begin{corollary}[Condition for perfect public recovery]
    There exists a sender $Q\in\mathcal{F}_R$ with $\Pr_Q[d(Z)=B]=1$ if and only if, for every $r$ with $\pi(r)>0$ and every $b$,
    \begin{equation}
        \sum_{z:d(z)=b}\beta_r(z)=p_r(b).
        \label{eq-perfect-recovery-partition-condition}
    \end{equation}
    \label{cor-perfect-public-recovery}
\end{corollary}

This is a mass-balance condition: within each task-relevant state, the reference probability assigned to messages decoded as $b$ must equal the probability of attribute value $b$. Neither a secret decoder nor a change to the receiver is required. Appendix~\ref{app-proof-perfect-public-recovery} proves both directions and constructs a witness sender. A sender that may change each task-conditioned message law slightly can do better.
\begin{corollary}[Recovery under bounded distortion]
    \label{cor-distortion-recovery}
    Let $d$ be onto. Over senders with $\mathrm{TV}(P_Q^{Z\mid r},\beta_r)\le\delta_r$ for every $r$, the largest value of $\Pr_Q[d(Z)=B]$ is $\sum_r\pi(r)[1-\max\{0,\mathrm{TV}(p_r,q_r^d)-\delta_r\}]$.
\end{corollary}
Perfect recovery is thus possible if and only if $\delta_r\ge\mathrm{TV}(p_r,q_r^d)$ for every $r$ (proof in Appendix~\ref{app-proof-perfect-public-recovery}).

\subsection{A Concrete Construction}\label{subsec-concrete-construction}

Consider $S=(R,B)$, with $R$ uniform on $16$ routes and $B$ an independent fair bit. Let $Z=(\widehat R,J)$, with $\widehat R=R$, and let the fixed receiver output $A=(\widehat R,A_2)=Z$, so $A_2=J$. The declared utility is $u(R,A)=\mathbf{1}\{\widehat R=R\}$.

The reference sender copies $R$ and draws $J$ as an independent fair bit. The alternative sender also copies $R$, but sets $J=B$. Both produce $\Pr[Z=(r,j)\mid R=r]=1/2$ for $j\in\{0,1\}$. Thus, they have exactly the same $(R,Z,A)$ law and task utility, even though the public decoder $d(\widehat r,j)=j$ recovers $B$ perfectly only under the alternative.

\begin{table}[t]
    \centering
    \caption{Exact witness. The reference and alternative have the same audited task law but different private-state dependence.}
    \label{tab-exact-witness}
    \begin{tabular}{@{}lcc@{}}
        \toprule
        Quantity & Reference & Alternative \\
        \midrule
        $(R,Z,A)$ law & $\lambda_0\rho$ & $\lambda_0\rho$ \\
        $I(S;Z)$ & $4$ bits & $5$ bits \\
        $I(B;Z\mid R)$ & $0$ bits & $1$ bit \\
        $U$ & $1$ & $1$ \\
        $\Pr[A_2=1]$ & $1/2$ & $1/2$ \\
        $D_B$ & $0$ & $1$ \\
        \bottomrule
    \end{tabular}
\end{table}

The receiver's second output can also have a different relationship to the private attribute. Define $D_B:=\Pr[A_2=1\mid B=1]-\Pr[A_2=1\mid B=0]$. This difference is $0$ for the reference and $1$ for the alternative, although $\Pr[A_2=1]=1/2$ under both. Table~\ref{tab-exact-witness} summarises the construction. It realises two couplings from Theorem~\ref{thm-task-preserving-coupling}: independence between $B$ and $J$, and perfect alignment $J=B$.

\subsection{What a Stronger Audit Fixes}\label{subsec-stronger-audit}

The construction also makes the role of the audit boundary clear. Observing $A$ in addition to $Z$ adds nothing when the receiver is fixed, as Sect.~\ref{sec-leakage-identification} showed. Even if the monitor receives a trusted $R$ alongside every message, the two senders above remain indistinguishable: their complete $(R,Z,A)$ distributions agree. In general, fixing this stronger law leaves the leakage range $[I_R,I_R+C_R]$, as established in Corollary~\ref{cor-task-preserving-leakage}.

Trusted information about the remaining state changes the situation. In the construction, an auditor given the true $B$ alongside $Z$ can distinguish the two senders: the total-variation distance between their $(B,Z)$ laws is $1/2$, giving optimal equal-prior discrimination success $3/4$ from a single paired observation. If the auditor instead receives a trusted pairing of the full private state $S$ and message $Z$, exact knowledge of their joint law determines $I_Q(S;Z)$ itself, though not what a receiver that shares a key with the sender learns (Sect.~\ref{subsec-selection-detectability}).

These statements concern the unrestricted sender class; Sect.~\ref{sec-selection-channels} analyses a concrete mechanism and a sender that attains the recovery optimum.

\section{Selection Channels in a Language-Model Sender}\label{sec-selection-channels}

We now consider a concrete mechanism available to an LLM sender, which generates several ordinary responses and chooses which one to emit. Such a sender is in general not in $\mathcal{F}_R$, and we characterise how far it departs from the reference law. By a \emph{selection channel} we mean the private-state-dependent choice among complete reference-generated messages, which steganography calls cover selection, not the shared-pad selection of cover positions described by Anderson and Petitcolas~\cite{anderson1998limits}.

\subsection{Selection Among Honest Samples}\label{subsec-selection-mechanism}

Let $B=g(S)$ be a binary private attribute and let $d:\mathcal{Z}\to\{0,1\}$ be a fixed public decoder. Given private state $s$, write $b_s=g(s)$ and
\begin{equation}
    a_s:=\Pr_0[d(Z)=b_s\mid S=s]
    \label{eq-honest-match-probability}
\end{equation}
for the probability that an ordinary reference message already decodes to the intended bit.

The selection sender $Q_k$ draws $k$ independent candidates from $P_0(\cdot\:|\:s)$ and emits the first candidate satisfying $d(Z)=b_s$. If none matches, it emits one additional, independent reference draw as a fallback. Every emitted message is therefore one of the samples produced by the reference procedure; the choice of which sample to emit is state-dependent. This is a bounded form of rejection sampling, closely related to the steganographic encoder of Hopper et al.~\cite{hopper2002provably} and to algorithm-substitution attacks that resample coins until a keyed function of the ciphertext reveals a key bit~\cite{bellare2014mass,bellare2015surveillance}. It requires control of the sender's sampling loop, as in a compromised agent runtime. Here the decoder is public and unkeyed, and the target bit may already be correlated with honest messages.

For any message $z$, this rule induces the kernel
\begin{equation}
\begin{aligned}
    Q_k(z\mid s)=P_0(z\mid s)\bigg[&(1-a_s)^k\\
      &+\mathbf{1}\{d(z)=b_s\}
        \sum_{i=0}^{k-1}(1-a_s)^i\bigg].
\end{aligned}
\label{eq-selection-kernel}
\end{equation}
The first term accounts for the fallback; the sum accounts for selecting the first matching candidate. In particular, the decoder's success probability is
\begin{equation}
    \Pr_{Q_k}[d(Z)=B\mid S=s]
    =1-(1-a_s)^{k+1}.
    \label{eq-selection-recovery-law}
\end{equation}
The extra exponent comes from the fresh fallback. Selection thus makes a private attribute reliably recoverable without changing the reference generator or the public decoder, but because matching candidates are preferred, $Q_k$ need not preserve $\beta$ or the task-conditioned laws $\beta_r$.

\subsection{Detectability}\label{subsec-selection-detectability}

The changed selection probabilities give a monitor that knows the mechanism a signal. Let $w_k(s,z)$ denote the bracketed factor in Eq.~\eqref{eq-selection-kernel}. On the support of the reference message law, the likelihood ratio between selected and reference messages is
\begin{equation}
    \frac{\bar Q_k(z)}{\beta(z)}
    =\mathbb{E}_0[w_k(S,z)\mid Z=z].
    \label{eq-selection-likelihood-ratio}
\end{equation}
It gives a message-only test without revealing the realised private state. A monitor with trusted $R$ can instead use the corresponding ratio conditional on $R=r$.

In a balanced case, detection depends only on information already present in honest messages. Write $\delta_k:=1-2^{-k}$.
\begin{theorem}[Detectability of selection, balanced case]\label{thm-selection-detectability}
Suppose $B$ is fair given each $R=r$ and $a_s=1/2$ for every state, and for $\beta_r(z)>0$ put $m_r(z):=\Pr_0[B=d(z)\mid Z=z,R=r]$. Within each route, selected and reference messages have likelihood ratio
\begin{equation}
    \ell_{k,r}(z)=1+2\delta_k\bigl(m_r(z)-\tfrac12\bigr),
    \label{eq-balanced-selection-ratio}
\end{equation}
and, with $D_r:=P_0^{(B,Z)\mid r}$ the honest joint law and $\Pi_r:=p_r\otimes\beta_r$,
\begin{equation}
\begin{aligned}
    \operatorname{TV}(P_{Q_k}^{Z\mid r},\beta_r)&=\delta_k\operatorname{TV}(D_r,\Pi_r),\\
    \chi^2(P_{Q_k}^{Z\mid r}\,\|\,\beta_r)&=\delta_k^2\,\chi^2(D_r\,\|\,\Pi_r).
\end{aligned}
\label{eq-selection-tv-identity}
\end{equation}
Over $T$ independent rounds with $R=r$, the largest advantage $\mathrm{TPR}-\mathrm{FPR}$ of any test lies between $1-e^{-Tc_r/12}$ and $\tfrac12(e^{Tc_r}-1)^{1/2}$, where $c_r:=\chi^2(P_{Q_k}^{Z\mid r}\|\beta_r)$. If $B$ and $Z$ are independent given $R$ under the reference, then $Q_k\in\mathcal{F}_R$ and the public decoder recovers $B$ with probability $1-2^{-(k+1)}$.
\end{theorem}
The detectable change is thus exactly proportional to how much honest messages already depend on $B$, and a monitor needs on the order of $1/c_r$ messages (proof in Appendix~\ref{app-proof-selection-canonical}). In the balanced case a more discreet honest sender therefore makes selection harder to detect, and one whose messages are independent of $B$ given $R$ makes it undetectable by any test of the task-relevant law. For unequal $a_s$, Eq.~\eqref{eq-selection-likelihood-ratio} is the general expression.

\noindent\textbf{Keyed senders.} If the sender selects on $d(z)=b_s\oplus\xi$ for a fresh uniform key bit $\xi$ per message, shared with the receiver and independent of $S$, then in the balanced case the emitted law is $P_0(\cdot\mid s)$ for every $s$, so no audit of $(S,Z,A)$ detects it, even with trusted pairings, while the receiver recovers $B$ with probability $1-2^{-(k+1)}$. Leakage to that receiver is $I(S;Z,\xi)$ rather than the unchanged $I_Q(S;Z)$, and Proposition~\ref{prop-canonical-balanced} still bounds it at a canonical interface.

\noindent\textbf{Law-preserving resampling within a route.} The bounded selection sender $Q_k$ generates candidates for the actual private state and generally changes its emitted message law. By contrast, an idealised resampling sender, with sampling access to the route's state distribution and to the reference sender, attains the optimal recovery of Corollary~\ref{cor-optimal-public-recovery} while preserving the full task-relevant message law. 

\begin{proposition}[Law-preserving resampling sender]\label{prop-resampling-sender}
Fix a public decoder $d$ and, for each $r$ with $\pi(r)>0$, a maximal coupling $\kappa_r^\star\in\Gamma(p_r,q_r^d)$. Assume sampling access to $\alpha_r$ and to the reference kernel $P_0(\cdot\mid r,n')$. On input $(r,n)$ with $b=g(r,n)$, draw a decoder target $Y=y$ with probability $\kappa_r^\star(b,y)/p_r(b)$. Then independently draw a fresh $n'\sim\alpha_r$ and $Z'\sim P_0(\cdot\mid r,n')$ on each trial until $d(Z')=y$, and emit the first accepted message. This procedure terminates almost surely, belongs to $\mathcal F_R$, and attains the maximum public-decoder recovery in Corollary~\ref{cor-optimal-public-recovery}.
\end{proposition}

Conditional on the selected target $y$, the accepted message has law $\beta_r$ conditioned on $d(Z)=y$. The resulting $(B,Z)$ law is precisely the lifted maximal coupling from Appendix~\ref{app-proof-public-recovery}; the sampling proof is in Appendix~\ref{app-resampling-proof}. With a binary decoder whose values both have positive probability, the sender needs two reference draws per message on average. Since the induced $(R,Z,A)$ law equals the reference law, no single-round test confined to those observations can distinguish the senders. Using $\widehat q_r^d$ in place of $q_r^d$ puts its law in route $r$ at total-variation distance exactly $\mathrm{TV}(\widehat q_r^d,q_r^d)$ from $\beta_r$; run this way on our samples, it recovers $B$ in $97.72\%$ of rounds while every monitor without the true $B$ stays near chance (Sect.~\ref{subsec-empirical-monitors}). If the reference never sees $N$, the sender's own draws already have law $\beta_r$, so no samples of other private states are needed. Applied round by round to an attribute that persists, as in our motivating example, both $Q_k$ and this sender repeat the decoded bit, which a monitor that knows $d$ notices within a few rounds; such an attribute can still leak without changing the transcript law (Sect.~\ref{subsec-composition-rounds}) if it is encoded in one round only and the sender is honest otherwise, provided honest messages depend on the private state only through $R$ and the attribute is independent of the task states. 

\subsection{Canonical Interfaces}\label{subsec-canonical-interfaces}

One way to restrict the communication channel is to enforce a canonical task interface. Let $C=c(Z)$ be a deterministic transformation with at most $K$ possible outputs. Then,
\begin{equation}
    I_Q(S;C)\leq H_Q(C)\leq\log_2 K.
    \label{eq-canonical-alphabet-bound}
\end{equation}
The same bound applies to $I_Q(B;C\mid R)$. It limits the information available after enforcement.

Suppose, moreover, that the receiver acts only through $c(Z)$, so that $\rho(a\mid z)=\widetilde\rho(a\mid c(z))$, and that canonicalisation is idempotent: $c(c(z))=c(z)$. Thus, the receiver also parses a canonical message to the same result, and replacing $z$ with $c(z)$ preserves the receiver's action law for every input message.

A bounded output alphabet does not by itself prevent leakage about $N$. A stronger guarantee is available when the canonical output nearly equals a task-only representation. Let $h$ map each task state to one of the $K\geq 2$ canonical outputs and suppose $\Pr_Q[C\neq h(R)]=\varepsilon$. Then, for any sender and any private attribute $B=g(R,N)$,
\begin{equation}
\begin{aligned}
    I_Q(B;C\mid R)&\leq I_Q(N;C\mid R)\leq H_Q(C\mid R)\\
    &\leq h_2(\varepsilon)+\varepsilon\log_2(K-1),
\end{aligned}
\label{eq-approx-canonical-leakage-bound}
\end{equation}
where $h_2$ is binary entropy. Indeed, conditional on $R$, the indicator of $C\neq h(R)$ is binary; if it is zero, $C$ is determined, and otherwise at most $K-1$ outputs remain. In the ideal case $\varepsilon=0$, we have $C=h(R)$ and hence $I_Q(N;C\mid R)=0$. The reference laws of the \emph{enforced} channel $C$ are then point masses, so $C_R$ of Eq.~\eqref{eq-residual-leakage-capacity}, computed with $C$ in place of $Z$, is zero.
\begin{proposition}[Balanced attribute and keys]\label{prop-canonical-balanced}
If, in addition, $B$ is fair given each $R=r$, then for every sender every decoder of $(C,R)$ recovers $B$ with probability within $\varepsilon$ of $1/2$, and $I_Q(B;C\mid R)\le2\varepsilon$. These bounds and Eq.~\eqref{eq-approx-canonical-leakage-bound} also hold with $(C,\xi)$ in place of $C$, for any key $\xi$ independent of $S$ shared by sender and decoder.
\end{proposition}

Alphabet restriction thus bounds total leakage, and agreement with a task-only representation bounds conditional leakage about the remaining private state for every decoder (derivations in Appendix~\ref{app-proof-selection-canonical}).

\section{Experimental Evaluation}\label{sec-empirical-evaluation}
We evaluate two senders with one open-weight model on one synthetic routing task, the selection sender $Q_k$, which changes the message law, and, in a post hoc arm, the law-preserving sender of Proposition~\ref{prop-resampling-sender}. We measure recovery against Eq.~\eqref{eq-selection-recovery-law}, the effect on routing, what generic and selection-aware monitors detect, the honest sender's own dependence on the attribute, and the residual information after canonicalisation.

\subsection{Setup}\label{subsec-empirical-setup}

We use $64$ equally weighted synthetic incident records with private state $S=(R,N)$. The task-relevant component $R$ specifies one of $16$ routes, formed by four fault domains and four severity levels; $N$ specifies one of four administrative variants. We target the private bit $B=N\bmod 2$. Each route contains two records with $B=0$ and two with $B=1$, so $H(B\mid R)=1$ bit. The honest prompt requests three lines giving a domain, severity, and short summary, while explicitly prohibiting disclosure of the administrative fields. The receiver is a fixed parser that extracts valid domain and severity fields and outputs one of the $16$ routes or \texttt{REJECT}. Routing is correct if that output equals the record's true route.

The sender is Qwen2.5-1.5B-Instruct~\cite{yang2024qwen25} at a fixed checkpoint revision. Generation uses temperature $1$ with no top-$k$, top-$p$, or repetition filtering: at each step, the next token is sampled from the model's unmodified next-token probability distribution, with a maximum of $96$ generated tokens. We write $W$ for the generated token sequence and $V$ for its rendered UTF-8 text. The public decoder $J=d(V)$ returns the low bit of the final SHA-256 digest byte of the exact rendered text. Both this decoder and the task receiver act on $V$; token-level likelihood calculations use $W$. In the empirical setting, the emitted message $Z$ of the formal model is represented by $V$.

As in Sect.~\ref{subsec-selection-mechanism}, $Q_k$ selects the earliest candidate whose decoded bit equals $B$ among $k\in\{1,2,4,8,16\}$ honest-model draws, or emits an independent fallback if none matches. Each round generates a single pool of $16$ candidates: the smaller-budget arms reuse prefixes of that pool and share the same fallback. Two further, independent honest draws, $\mathsf{H}$ and $\mathsf{H}'$, are excluded from selection. The shared pool makes the $Q_k$ arms paired, rather than independent experimental samples.

Primary results use $8{,}192$ held-out rounds per arm ($64$ states, eight base seeds, and $16$ rounds per state and seed). Separate training and threshold-calibration splits contain $1{,}024$ rounds each; an independent $4{,}096$-draw stream calibrates the honest decoder-match probabilities used in the selection-law predictions. Distinct random-number streams are derived for each data split, task state, round, and draw role. Intervals use $5{,}000$ paired bootstrap resamples of complete round bundles; seeds, stratification, compute and further details are in Appendix~\ref{app-experiments}.

\subsection{Recovery and Task Accuracy}\label{subsec-empirical-recovery-task}

\begin{figure*}[t]
    \centering
    \includegraphics[width=0.85\textwidth]{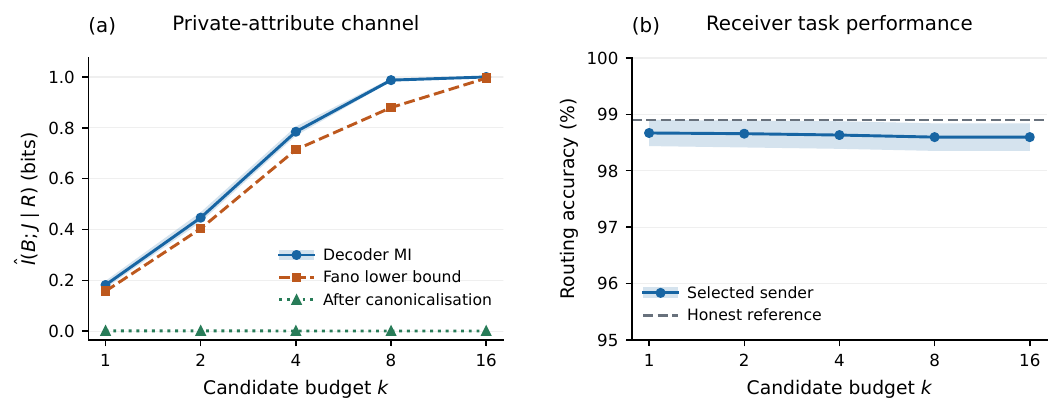}
    \caption{Candidate selection increases conditional private-attribute information while routing accuracy remains high. (a)~Empirical $\widehat I(B;J\mid R)$, pointwise $95\%$ lower confidence bounds on $I(B;V\mid R)$ from decoder error (Appendix~\ref{app-exp-statistics}), and $\widehat I(B;J\mid R)$ after canonicalisation. (b)~Routing accuracy of selected messages and the honest reference. Bands are $95\%$ bootstrap intervals; $k$ is on a $\log_2$ scale.}
    \label{fig-information-task-budget}
\end{figure*}

Fig.~\ref{fig-information-task-budget} shows the relationship between candidate budget, private-attribute information, and task performance. Since $J$ is a deterministic function of the emitted text, data processing gives $I_Q(B;V\mid R)\geq I_Q(B;J\mid R)$. We estimate the latter from the empirical $(B,J)$ distribution separately within each route. The resulting $\widehat I(B;J\mid R)$ increases from $0.1817$ bits at $k=1$ to $0.4465$, $0.7845$, and $0.9876$ bits at $k=2,4,8$, reaching $1.0000$ bit at $k=16$. The corresponding public-decoder recovery rises from $74.41\%$ to $100.00\%$; under the honest sender $\mathsf{H}$, recovery is $49.90\%$ and the empirical decoder information is $0.0028$ bits.

Independent calibration finds that the reference hash bit is approximately balanced within every route: $\widehat q_r^d(1)$ ranges from $0.449$ to $0.543$. Substituting these estimates into Corollary~\ref{cor-optimal-public-recovery} gives a plug-in maximum recovery of about $98.0\%$ for law-preserving senders such as that of Proposition~\ref{prop-resampling-sender}, biased downwards by sampling error. By Fano's inequality, they also give a plug-in lower bound $C_B\ge 1-h_2(0.020)\approx 0.86$ bits, so fixing the task-relevant law leaves most of the private bit unconstrained.

Eq.~\eqref{eq-selection-recovery-law} predicts recovery $1-(1-a_s)^{k+1}$ for each private state, including the fresh fallback. Using match probabilities from the independent calibration stream ($64$ draws per state) and averaging over the $64$ states gives predicted recoveries of $74.96\%$, $87.13\%$, $96.44\%$, $99.68\%$, and $100.00\%$ for the five budgets. The corresponding observed values are $74.41\%$, $87.02\%$, $96.51\%$, $99.87\%$, and $100.00\%$; the largest difference is $0.55$ percentage points. 

At $k=16$, the decoder matches $B$ in all $8{,}192$ test rounds. Writing $\eta=\Pr[J\neq B]$, binary Fano's inequality gives $I_{Q_k}(B;V\mid R)\geq 1-h_2(\eta)$ with $h_2$ the binary entropy~\cite{cover2006elements}, and the one-sided zero-error Clopper--Pearson bound~\cite{clopper1934use}, which also covers the design-averaged error of independent non-identical trials (Appendix~\ref{app-exp-statistics}), yields $\eta\leq 1-0.05^{1/8192}=3.66\times10^{-4}$ at $95\%$ confidence, and hence $I_{Q_{16}}(B;V\mid R)\geq0.9953$ bits.

The legitimate task remains reliable across the same budget range. The honest arm $\mathsf{H}$ routes $8{,}102$ of $8{,}192$ incidents correctly ($98.90\%$) and an independent honest replicate $\mathsf{H}'$ routes $8{,}085$ ($98.69\%$), while $Q_{16}$ routes $8{,}077$ correctly ($98.60\%$), a paired change of $-0.31$ percentage points ($95\%$ bootstrap interval $[-0.65,0.04]$). Accuracy alone does not compare the full receiver behaviour, so we also estimate the route-conditional distribution over all $16$ receiver outputs plus rejection. Its average total-variation distance from $\mathsf{H}$ is $0.004517$ for $Q_{16}$, and independent honest samples $\mathsf{H}'$ differ from $\mathsf{H}$ by the same amount (both are $74/16{,}384$), the level of honest replication.

Although the public hash decoder is at chance under $\mathsf{H}$, honest messages are not free of attribute information. From honest token-sequence likelihoods we estimate $I_0(B;W\mid R)=0.2620$ bits under the suppression instruction, against $0.7368$ bits for an honest control whose prompt asks for the fields; since $V$ is a function of $W$, the rendered text carries at most this much. The model does not fully comply, and $18.8\%$ of honest messages state at least one administrative value ($61.1\%$ when the prompt asks for them). The instruction thus reduces but does not remove the honest dependence on $B$, which is what a selection-aware monitor can exploit (Sect.~\ref{subsec-empirical-monitors}).

\begin{table}[hb!]
    \centering
    \caption{Detection of $Q_{16}$ versus $\mathsf{H}$, grouped by monitor access; AUC$^\ast$ is for the law-preserving sender (post hoc arm). $n$ is per arm; (1)/(16) gives messages per score. Intervals are in Table~\ref{tab-app-auc-intervals}.}
    \label{tab-empirical-monitors}
    \footnotesize
    \setlength{\tabcolsep}{3pt}
    \begin{tabular}{@{}lrrrrr@{}}
        \toprule
        Monitor & $n$ & FPR & TPR & AUC & AUC$^\ast$ \\
        \midrule
        \multicolumn{6}{@{}l}{\emph{No trusted private-state information}} \\
        Structure (1) & 8,192 & 3.91 & 4.10 & .504 & .503 \\
        Character text (1) & 8,192 & 6.35 & 6.23 & .490 & .490 \\
        Surprisal (1) & 1,024 & 14.06 & 14.26 & .496 & .495 \\
        Selection-aware ratio (1) & 1,024 & 4.00 & 8.30 & .645 & .504 \\
        Selection-aware ratio (16) & 64 & 1.56 & 56.25 & .972 & .540 \\
        \addlinespace[2pt]
        \multicolumn{6}{@{}l}{\emph{Trusted task route $R$ (identical to the rows above)}} \\
        Route surprisal (1) & 1,024 & 14.06 & 14.26 & .496 & .495 \\
        Route-aware ratio (1) & 1,024 & 4.00 & 8.30 & .645 & .504 \\
        Route-aware ratio (16) & 64 & 1.56 & 56.25 & .972 & .540 \\
        \addlinespace[2pt]
        \multicolumn{6}{@{}l}{\emph{Trusted private attribute $B$}} \\
        Known-code control (16) & 512 & 2.93 & 100.00 & 1.000 & 1.000 \\
        \bottomrule
    \end{tabular}
\end{table}


\subsection{Monitors}\label{subsec-empirical-monitors}

We next test whether a passive monitor can distinguish selected responses $Q_{16}$ from independent honest responses $\mathsf{H}$. The message-only monitors use either structural features of the rendered text $V$, character-level text features, or reference-model surprisal computed from the retained token sequences $W$. A selection-aware monitor instead uses the likelihood ratio derived from the known selection rule, reference state-conditioned likelihoods, and independently calibrated match probabilities (Sect.~\ref{subsec-selection-detectability}). We also evaluate variants given a trusted task route $R$, and a stronger positive control given the private bit $B$ and the public code. Scoring $W$ rather than $V$, under all $64$ records, gives the likelihood-based monitors more than a deployed passive monitor would have, so their results are optimistic for the monitor; only the positive control receives the realised $B$.

Learned classifiers are fitted on training observations; nominal $5\%$ false-positive thresholds are set on separate honest calibration observations and then held fixed for testing, with separate calibration for collection-level scores. Table~\ref{tab-empirical-monitors} reports the actual held-out false-positive rate (FPR), true-positive rate (TPR), and area under the ROC curve (AUC), with $n$ observations per arm; $(1)$ denotes single messages and $(16)$ denotes collections of $16$ messages. The surprisal threshold was calibrated on a novelty-purged cohort, so its FPR on the full test stream is $14.06\%$ (Appendix~\ref{app-exp-monitors}).

The generic single-message monitors discriminate poorly: their AUCs are near $0.5$. In contrast, the selection-aware ratio gives an AUC of $0.645$ on individual messages and $0.972$ when its scores are aggregated over $16$ messages. At its calibrated collection threshold, the latter detects $56.25\%$ of attack collections with $1.56\%$ FPR. Supplying trusted $R$ leaves every score unchanged to the printed precision, because almost every message states its route, so conditioning on $R$ shifts surprisal by a nearly constant $4$ bits. By comparison, the trusted-$B$ known-code positive control attains AUC $1.000$ and $100\%$ TPR over collections, but assumes information unavailable at the passive boundary.

Selection thus creates a distributional signal although every candidate comes from the honest generator, and the weak performance of generic monitors does not mean that $Q_{16}$ preserves the honest message law. Theorem~\ref{thm-selection-detectability} predicts its size from honest messages alone. In this approximately balanced design the honest posteriors give a total-variation distance of $(1-2^{-16})\,\mathbb{E}_0|\Pr_0[B=1\mid W,R]-1/2|=0.189$ between $Q_{16}$ and $\mathsf{H}$, so no single-message monitor can exceed AUC $1/2+0.189-0.189^2/2=0.671$; the selection-aware ratio reaches $0.645$ ($95\%$ interval $[0.622,0.667]$). Explicit administrative statements explain at most about half of this distance, even if each revealed $B$ completely. 

\noindent\textbf{A law-preserving sender.} A post hoc arm runs the sender of Proposition~\ref{prop-resampling-sender} on the stored test candidates, without new generation; each trial draws a fresh record of the same route, which is the sampling access the proposition assumes. It recovers $B$ in $97.72\%$ of rounds, against the $98.00\%$ that Corollary~\ref{cor-optimal-public-recovery} predicts from the calibration estimates, with $2.03$ draws per message and routing accuracy $98.95\%$ ($98.90\%$ for $\mathsf{H}$). Every monitor without the true $B$, applied unchanged, stays near chance (Table~\ref{tab-empirical-monitors}, AUC$^\ast$), including the selection-aware ratio, which gives AUC $0.540$ over $16$ messages against $0.972$ for $Q_{16}$. Only the known-code control, given the true $B$, still detects the channel, as Sect.~\ref{subsec-stronger-audit} predicts. By Corollary~\ref{cor-distortion-recovery}, moving each route law by only $\mathrm{TV}(p_r,q_r^d)$, about $0.02$ on average in plug-in estimates, would lift recovery to $100\%$.

\subsection{Canonical Interface}\label{subsec-empirical-canonicalization}

Finally, we restrict communication to exactly the fields used by the legitimate receiver. The deterministic canonicaliser parses the emitted text $V$ using the receiver's route parser and emits one fixed serialisation for each of the $16$ accepted routes, or the single string \texttt{REJECT}. Write $A$ for the parsed action and $C=c(V)$ for the enforced message. Thus $C$ is a function of $A$, takes at most $K=17$ values, and requires no access to the private record or its selected bit. Parsing the canonical string reproduces $A$, so canonicalisation is idempotent and preserves each receiver decision by construction, including incorrect routes and rejections.

For $Q_{16}$, recovery by the unchanged public hash decoder falls from $100.00\%$ before enforcement to $49.95\%$ afterwards. The empirical conditional information $\widehat I(B;J\mid R)$ falls from $1.0000$ to $0.000629$ bits, a paired reduction of $0.9994$ bits. Fig.~\ref{fig-information-task-budget} shows the same return towards chance recovery across candidate budgets. As an implementation check, the parsed action agrees before and after canonicalisation in all $57{,}344$ test comparisons across the seven arms, so $Q_{16}$ keeps its $98.60\%$ routing accuracy.

The near-chance recovery is also constrained by the interface, rather than being peculiar to the chosen hash decoder. In the balanced evaluation population, $B$ is a fair bit conditional on each true route $R$. Since $C=h(R)$ whenever routing is correct, any binary decoder of the canonical message, even one additionally given $R$, agrees with $B$ with probability $1/2$ on the corresponding task-only representation and can differ from this baseline only on routing errors. For $Q_{16}$, the observed error fraction is $\widehat\varepsilon=115/8192=1.40\%$, so every decoder has test-population recovery in $[48.60\%,51.40\%]$. Moreover, applying Eq.~\eqref{eq-approx-canonical-leakage-bound} to the empirical joint distribution gives
\begin{equation}
\begin{aligned}
    \widehat I(B;C\mid R)
    &\leq h_2(\widehat\varepsilon)+\widehat\varepsilon\log_2(16)\\
    &\approx 0.163\ \text{bits}.
\end{aligned}
\label{eq-empirical-canonical-leakage-bound}
\end{equation}
This bounds the conditional information in the whole canonical message, and hence in every decoder of it, at the observed routing error (Appendix~B.6). Because $B$ is fair given each route, Proposition~\ref{prop-canonical-balanced} sharpens it to $\widehat I(B;C\mid R)\le2\widehat\varepsilon\approx0.028$ bits; with the one-sided $95\%$ Clopper--Pearson bound $\varepsilon\le0.0164$, the bound is $0.033$ bits.

Unlike the monitoring results above, this bound holds for every sender behind a trusted canonicaliser, including one that shares a key with the receiver, in terms of that sender's routing-error rate, so a sender can leak more only by misrouting more. Receivers that need free text are outside its scope.

\section{Related Work}\label{sec-related-work}

\noindent\textbf{Multi-agent security and secret collusion.} Communication between AI agents creates opportunities for unwanted information exchange that are not captured by studying agents in isolation~\cite{dewitt2025openchallenges,hagag2026architecture}. Motwani et al.~\cite{motwani2024secret} formalise secret collusion among generative agents, including the roles of communication channels, monitoring, common knowledge, and incentives; they show that steganography can be detected with grounding functions, which need the agent's knowledge, and bound what ideal paraphrasing leaves by semantic entropy. Our monitor lacks that knowledge, and Eq.~\eqref{eq-approx-canonical-leakage-bound} is the counterpart of their bound for a canonical task interface, in terms of task error. Their earlier benchmark examines collusion using steganographic tools~\cite{motwani2023perfect}, while subsequent work studies steganographic communication emerging from optimisation and its resistance to monitoring or paraphrasing~\cite{mathew2025hidden}. Related investigations address hidden reasoning within individual models~\cite{roger2023hiding} and covert communication enabled by tool use~\cite{rippin2026tooluse}. These studies motivate treating inter-agent messages as a security boundary, but largely investigate the construction, emergence, or detection of particular behaviours. We instead ask what agreement of the audited communication distribution can certify about the sender's dependence on private state, irrespective of how an alternative sender is implemented.

\noindent\textbf{Steganographic security and monitor strength.} Provably secure steganography seeks to transmit hidden content without changing the distribution of ordinary covertext, with information-theoretic~\cite{cachin1998information} or complexity-theoretic~\cite{hopper2002provably} security; Meteor~\cite{kaptchuk2021meteor} and Discop~\cite{ding2023discop} achieve this with language models, and Wang and Moulin~\cite{wang2008perfectly} derive its capacity with a shared key. More recently, Vaikuntanathan and Zamir~\cite{vaikuntanathan2026undetectable} construct covert conversations that are computationally indistinguishable from honest interaction under cryptographic assumptions, including against a strong passive auditor with knowledge of the agents' private contexts. That auditor is stronger than ours, but the guarantee is computational. Our main results are information-theoretic and keyless and characterise what a weaker passive audit identifies; with a fresh shared key bit per message, even a paired audit fails while the interface bounds survive (Sect.~\ref{subsec-selection-detectability}), and undetectable watermarks~\cite{christ2024undetectable} are keyed embeddings of this kind in LLM output. For imperfect stegosystems, detection evidence accumulates over many covers, as in the square-root law of steganographic capacity~\cite{ker2008square}; for selection, Theorem~\ref{thm-selection-detectability} gives the per-message $\chi^2$ divergence exactly and hence, up to a constant factor, the number of messages a monitor needs.

\noindent\textbf{Couplings and minimum entropy.} Particularly close to our mathematical approach, Schroeder de Witt et al.~\cite{dewitt2023perfect} show that perfectly secure steganographic procedures correspond to couplings and relate optimal coding efficiency to minimum-entropy couplings. The entropy properties of couplings themselves have been studied independently~\cite{kovacevic2015entropy}. With the private state in place of the hidden message, Proposition~\ref{thm-leakage-interval} is the keyless counterpart of this coupling view of perfect security~\cite{cachin1998information,dewitt2023perfect}, and the general point that an output law does not determine dependence on secrets is the probabilistic form of the refinement paradox~\cite{jacob1989derivation,morgan2006shadow}. The audit setting differs in two respects that our other results address. Honest messages may already depend on the private state, whereas a covertext is independent of the hidden message, and in a balanced case this dependence fixes how detectable selection is (Theorem~\ref{thm-selection-detectability}). The auditor also asks what the observed law identifies, which yields sharp leakage and recovery limits under task preservation and the condition under which passive conformance does determine leakage (Corollaries~\ref{cor-task-preserving-leakage} to~\ref{cor-distortion-recovery}).

\noindent\textbf{Confinement and quantitative information flow.} Lampson's confinement problem asks how to prevent a program, during its execution, from transmitting information to parties other than its caller~\cite{lampson1973confinement}. Noninterference and information-flow policies formalise whether secret inputs can influence public observations~\cite{sabelfeld2003language,clarkson2010hyperproperties}; for probabilistic systems the condition is stated on output laws conditioned on the secret~\cite{gray1991toward}, which requires exactly the pairing our monitor lacks. Quantitative information flow offers several ways to measure disclosure: Shannon mutual information quantifies statistical dependence, whereas min-entropy leakage captures changes in one-guess vulnerability~\cite{smith2009foundations}, and gain-function-based $g$-leakage extends this perspective to other adversarial objectives~\cite{alvim2012gain,alvim2020science}. Our main results use Shannon leakage, and Corollary~\ref{cor-other-measures} shows that the same interval structure holds for every multiplicative $g$-leakage, with an exact, though in general NP-hard, expression for Bayes vulnerability. These measures are ordinarily evaluated for a specified channel, whereas the deployed kernel $Q(z\mid s)$ is unavailable to our passive monitor. Strategic confinement additionally notes that small information capacity need not imply small downstream harm when agents target consequential predicates~\cite{dewitt2026notestrategicconfinementproblem}. Our attribute-specific leakage and recovery results identify which such predicates can remain exposed despite task-law preservation.

\noindent\textbf{Monitoring and estimating information flow.} Information-flow policies are hyperproperties~\cite{clarkson2010hyperproperties}, runtime monitors for them compare several executions~\cite{agrawal2016runtime,finkbeiner2017monitoring}, and execution monitors cannot enforce them~\cite{schneider2000enforceable}. For a randomised sender such as an LLM, only distributional statements are possible, and our results show that even the exact message distribution does not determine the leakage. Statistical leakage estimators~\cite{chothia2011statistical,cherubin2019fbleau} rely on secrets paired with outputs, which is the trusted pairing our monitor lacks. In algorithm-substitution attacks, a subverted implementation leaks through outputs distributed like honest ones~\cite{bellare2014mass}; watchdogs test it against its specification~\cite{russell2016cliptography}, like our monitor, and reverse firewalls~\cite{mironov2015reverse} and active wardens~\cite{fisk2002eliminating} rewrite its outputs, like our canonical interface.

\noindent\textbf{Partial identification from incomplete observations.} The distinction between knowing marginal distributions and knowing how variables are jointly paired also appears in statistical matching and partial identification~\cite{dorazio2006statistical,manski2003partial}. There, several joint distributions agree with what is observed, and the right conclusion is a set of compatible values rather than a point estimate. Our compatible leakage set applies this view to communication security, with trusted state--message pairings as the missing data and leakage as the parameter. Fixing the task-relevant law does not merely narrow an interval; it fixes a different family of marginals and leaves independent coupling choices within task states, which gives population-level leakage and recovery limits rather than a finite-sample estimator.

\noindent\textbf{Privacy--utility trade-offs and secure agent pipelines.} Information-theoretic privacy research designs release mechanisms that balance useful disclosure against statistical inference of private attributes~\cite{sankar2013utility,calmon2012privacy,makhdoumi2014privacy}. Related agent-security work develops information-flow controls for AI planners~\cite{costa2025informationflow}, bounds and privacy-regularised training for sequential multi-agent LLM pipelines~\cite{asif2026privacy}, and decision-theoretic measures of steganography based on differences in what downstream agents can infer or use~\cite{anwar2026decision}. The last avoids a reference law by comparing what a decoding receiver and a non-decoding sentinel gain from a signal; we instead give our monitor the reference law of $(R,Z)$, and with it every declared utility, and characterise what that knowledge alone certifies about leakage and recovery.

\section{Conclusion}\label{sec-conclusion}

Matching the authorised distribution of messages is not a privacy guarantee at a multi-agent communication boundary. For a single round we have shown how much it leaves open. Compatible leakage forms sharp intervals even when the entire task-relevant law and every declared utility are fixed, and total leakage is determined if and only if, in every task state, the remaining private state or the authorised message is determined by that state. For a sender that preserves the audited law, in independent rounds or for the whole transcript, more observations of the audited variables cannot recover the missing state--message dependence.

For a sender, such as an LLM, that selects among its own responses, the change to the message law is, in a balanced case, proportional to the honest sender's own dependence on the private attribute, so more discreet honest agents give this attack better cover, and fresh key bits shared with the receiver can hide it even from an audit with trusted pairings. In our experiment it recovers a private bit in every held-out round at $98.60\%$ routing accuracy, generic monitors stay near chance, and a selection-aware monitor detects it over collections of $16$ messages, whereas a law-preserving sender removes that signal at the predicted recovery. Conformance testing therefore cannot exclude a sender that preserves the audited law, however exactly that law is known, whereas a canonical interface preserves every receiver decision and bounds, in terms of its routing error, what any sender, keyed or not, reveals to a receiver that acts only on its output. Extending the analysis to adaptive senders and richer interactions is the natural next step.

\appendices

\section*{Acknowledgements}

We used large language models both as systems under evaluation and as assistants during the preparation of this work. In the experiments, Qwen2.5-1.5B-Instruct served as the reference language-model sender from which candidate messages were generated and selected. The evaluation used a fixed model revision and locally controlled inference; it was not a comparison across multiple model families.

Separately, we used Claude Opus 4.8 and ChatGPT 5.6 Sol to assist with implementation, debugging, analysis workflows, and editorial refinement of the manuscript. Their outputs were treated as suggestions rather than authoritative evidence and were reviewed by the authors. The authors take full responsibility for the mathematical arguments, experimental results, citations, and final content of this work.


\bibliographystyle{IEEEtran}
\bibliography{biblio}


\section*{Ethics and Responsible Research}

This work addresses a dual-use security problem: a sender may exploit legitimate inter-agent communication to disclose private information. Our exact constructions characterise such information flow under specified observation boundaries, and our LLM experiment evaluates one candidate-selection mechanism. These results could also inform attempts at covert disclosure. We therefore distinguish the formal indistinguishability statements from the empirical attack, which changes the emitted message law and can be detected by a selection-aware monitor.

We limited the experimental risks by using synthetic incident records, an artificial private attribute, and a locally executed language model in a controlled routing simulation. The study did not use real personal or medical records, involve human participants, compromise deployed agent systems, or send genuine private information to an external recipient. The local-to-cloud example motivating the work is a threat scenario, not an account of an attack conducted in the experiments.

The security benefit is to establish which leakage claims are unsupported by passive communication monitoring and to evaluate stronger safeguards. In particular, we report both the detectability of the selection mechanism and the reduction in private-attribute information afforded by a canonical task interface. We consider the value of making these audit limitations and defensive alternatives explicit, together with the controlled experimental design, to outweigh the residual risk of misuse.

\section{Proofs}\label{app-proofs}

\subsection{Fixed message law}\label{app-proof-fixed-law}
\begin{proof}[Proof of Proposition~\ref{thm-leakage-interval}]
For each sender kernel $Q$ satisfying $\bar Q=\beta$, define $\gamma_Q(s,z):=\mu(s)Q(z\mid s)$. Its state marginal is $\mu$, and its message marginal is $\bar Q=\beta$, so $\gamma_Q\in\Gamma(\mu,\beta)$. Conversely, for any $\gamma\in\Gamma(\mu,\beta)$, set $Q_\gamma(z\mid s):=\gamma(s,z)/\mu(s)$. This is well defined because $\mu(s)>0$ on the retained state space; it is a stochastic kernel because $\sum_z\gamma(s,z)=\mu(s)$, and its emitted message law is $\sum_s\mu(s)Q_\gamma(z\mid s)=\beta(z)$. Thus the feasible sender kernels and the couplings in $\Gamma(\mu,\beta)$ are in one-to-one correspondence, with $I_Q(S;Z)=I_{\gamma_Q}(S;Z)$.

The coupling set is nonempty, since it contains $\gamma_{\mathrm{ind}}:=\mu\otimes\beta$, and is a closed, bounded subset of the finite-dimensional probability simplex; hence it is compact. For any $\gamma\in\Gamma(\mu,\beta)$, the marginal entropies are fixed and $I_\gamma(S;Z)=H(\mu)+H(\beta)-H(\gamma)$. Mutual information is nonnegative, with $I_{\gamma_{\mathrm{ind}}}(S;Z)=0$. Since finite-alphabet entropy is continuous, there is a minimum-entropy coupling $\gamma_\star\in\Gamma(\mu,\beta)$, which attains $I_{\gamma_\star}(S;Z)=G(\mu,\beta)$. Consequently, every feasible leakage value lies in $[0,G(\mu,\beta)]$, and both endpoints are feasible.

To establish attainability of every intermediate value, for $t\in[0,1]$ let $\gamma_t:=(1-t)\gamma_{\mathrm{ind}}+t\gamma_\star$. The prescribed marginals are linear constraints, so $\gamma_t\in\Gamma(\mu,\beta)$ for all $t$. The map $t\mapsto I_{\gamma_t}(S;Z)$ is continuous, and its values at $t=0$ and $t=1$ are $0$ and $G(\mu,\beta)$, respectively. By the intermediate value theorem it attains every value between these endpoints. Applying the kernel construction above to each $\gamma_t$ proves $\mathcal I(\beta)=[0,G(\mu,\beta)]$, as claimed. By the lemma at the end of Appendix~\ref{app-proof-task-leakage}, $G(\mu,\beta)=0$ exactly when $\mu$ or $\beta$ is a point mass.
\end{proof}

\noindent\textit{Proof of Corollary~\ref{cor-other-measures}.} For a gain function $g:\mathcal{W}\times\mathcal{S}\to\mathbb{R}$ on a finite action set $\mathcal{W}$ with prior vulnerability $V_g(\mu):=\max_w\sum_s\mu(s)g(w,s)>0$, the posterior vulnerability of a coupling $\gamma\in\Gamma(\mu,\beta)$ is
\begin{equation*}
V_g(\gamma)=\sum_z\max_{w}\sum_s\gamma(s,z)\,g(w,s),
\end{equation*}
and its multiplicative $g$-leakage is $\log_2\bigl(V_g(\gamma)/V_g(\mu)\bigr)$. Each summand is a maximum of finitely many linear functions of $\gamma$, so $V_g$ is convex and continuous on $\Gamma(\mu,\beta)$. A sum of maxima is at least the maximum of the sum, so $V_g(\gamma)\ge\max_w\sum_{s,z}\gamma(s,z)g(w,s)=V_g(\mu)$, with equality at $\mu\otimes\beta$; the smallest compatible leakage is therefore $0$. A convex function on the polytope $\Gamma(\mu,\beta)$ attains its maximum at a vertex, and since $\Gamma(\mu,\beta)$ is convex and $V_g$ is continuous, every intermediate value is attained, as in the proof of Proposition~\ref{thm-leakage-interval}. For Bayes vulnerability, $\mathcal{W}=\mathcal{S}$ and $g(w,s)=\mathbf{1}\{w=s\}$, so $V(\gamma)=\sum_z\max_s\gamma(s,z)=\max_d\Pr_\gamma[d(Z)=S]$ over maps $d:\mathcal{Z}\to\mathcal{S}$, and $V(\mu)=\max_s\mu(s)$. Maximising over $\gamma$ and applying Corollary~\ref{cor-optimal-public-recovery} with a constant route and $B=S$ gives $\max_\gamma V(\gamma)=\max_d[1-\mathrm{TV}(\mu,q^d)]$.\hfill$\square$

Since $P_0$ itself has marginal $\beta$, its leakage $I_0(S;Z)$ belongs to this interval. Subtracting this fixed reference value from the interval proved above gives Eq.~\eqref{eq-compatible-increment-range}, the claimed range of reference-relative leakage increments.

\subsection{Task-conditioned attribute--message couplings}\label{app-proof-task-coupling}

\begin{proof}[Proof of Theorem~\ref{thm-task-preserving-coupling}]
Fix any $r$ with $\pi(r)>0$. For every $Q\in\mathcal{F}_R$, the law of the private state is fixed, so $P_Q(B=b\mid R=r)=p_r(b)$. By the definition of $\mathcal{F}_R$, we also have $P_Q(Z=z\mid R=r)=\beta_r(z)$. Hence $P_Q^{(B,Z)\mid R=r}$ has the prescribed marginals and belongs to $\Gamma(p_r,\beta_r)$. This proves the inclusion from left to right in Eq.~\eqref{eq-task-preserving-coupling-complete}.

For the reverse inclusion, choose an arbitrary coupling $\eta_r\in\Gamma(p_r,\beta_r)$ for \emph{each} $r$ with $\pi(r)>0$. Define a sender kernel on every state $(r,n)$ of positive probability by
\begin{equation}
    Q_\eta(z\mid r,n)
    :=\frac{\eta_r(g(r,n),z)}{p_r(g(r,n))}.
\end{equation}
The denominator is strictly positive on this support: if $\mu(r,n)=\pi(r)\alpha_r(n)>0$ and $b=g(r,n)$, then $p_r(b)\geq\alpha_r(n)>0$. Since $\eta_r$ is nonnegative and has first marginal $p_r$, we have $Q_\eta(z\mid r,n)\geq0$ and $\sum_z Q_\eta(z\mid r,n)=1$. Values of the kernel on zero-probability states, if retained, may be chosen arbitrarily.

For every $b$ with $p_r(b)>0$, the resulting conditional joint law satisfies
\begin{equation}
\begin{aligned}
 &P_{Q_\eta}(B=b,Z=z\mid R=r)\\
 &\quad=\sum_{n:g(r,n)=b}\alpha_r(n)Q_\eta(z\mid r,n)\\
 &\quad=\frac{\eta_r(b,z)}{p_r(b)}
        \sum_{n:g(r,n)=b}\alpha_r(n)\\
 &\quad=\eta_r(b,z).
\end{aligned}
\end{equation}
If $p_r(b)=0$, both sides are zero, since the event $B=b$ is impossible conditional on $R=r$ and the $B$-marginal of $\eta_r$ is $p_r$. Summing the displayed identity over $b$ now yields $P_{Q_\eta}(Z=z\mid R=r)=\sum_b\eta_r(b,z)=\beta_r(z)$. Therefore $P_{Q_\eta}^{(R,Z)}=\lambda_0$, so $Q_\eta\in\mathcal{F}_R$, and its conditional $(B,Z)$ law equals the prescribed $\eta_r$.

The construction specifies one kernel $Q_\eta(z\mid r,n)$ for every state in the support of $\mu$, using only the coupling chosen for that state's value of $r$. Thus the choices of $\eta_r$ impose no cross-route consistency constraints and are realised \emph{simultaneously} by this single sender. Since the family of couplings was arbitrary, the reverse inclusion and the final assertion of the theorem follow.
\end{proof}

\subsection{Sharp leakage intervals under task-law preservation}\label{app-proof-task-leakage}

\begin{proof}[Proof of Corollary~\ref{cor-task-preserving-leakage}]
Every $Q\in\mathcal F_R$ has $P_Q^{(R,Z)}=\lambda_0$, so $I_Q(R;Z)=I_0(R;Z)=I_R$. Since $S=(R,N)$, the chain rule and conditioning on $R$ give
\begin{equation}
\begin{aligned}
I_Q(S;Z)&=I_R+I_Q(N;Z\mid R),\\
I_Q(N;Z\mid R)&=\sum_{r:\pi(r)>0}\pi(r)I_Q(N;Z\mid R=r).
\end{aligned}
\end{equation}
For each such $r$, the conditional joint law of $(N,Z)$ has marginals $\alpha_r$ and $\beta_r$. Therefore, by the definition of $G$, $0\leq I_Q(N;Z\mid R=r)\leq G(\alpha_r,\beta_r)$, and consequently $I_R\leq I_Q(S;Z)\leq I_R+C_R$.

To show that these endpoints and every intermediate value are attainable, fix, for each $r$ with $\pi(r)>0$, a maximiser $\gamma_r^\star\in\Gamma(\alpha_r,\beta_r)$, which exists by compactness and continuity on finite alphabets. Let $\gamma_r^0:=\alpha_r\otimes\beta_r$ and, for $t\in[0,1]$, put $\gamma_r^t:=(1-t)\gamma_r^0+t\gamma_r^\star$. Each $\gamma_r^t$ has the prescribed marginals. Applying Theorem~\ref{thm-task-preserving-coupling} with the attribute $B=N$, the family $(\gamma_r^t)_r$ is realised by a single sender $Q_t\in\mathcal F_R$. Its total leakage is
\begin{equation}
I_{Q_t}(S;Z)=I_R+\sum_r\pi(r)I_{\gamma_r^t}(N;Z).
\end{equation}
This quantity varies continuously with $t$, equals $I_R$ at $t=0$, and equals $I_R+C_R$ at $t=1$. The intermediate value theorem thus gives the entire interval $[I_R,I_R+C_R]$, including both endpoints.

For the attribute-specific range, Theorem~\ref{thm-task-preserving-coupling} likewise identifies the feasible conditional laws of $(B,Z)$ with $\Gamma(p_r,\beta_r)$, independently for every $r$. Hence
\begin{equation}
0\leq I_Q(B;Z\mid R)=\sum_r\pi(r)I_Q(B;Z\mid R=r)\leq C_B.
\end{equation}
Choose, for each positive-probability $r$, a maximiser $\eta_r^\star\in\Gamma(p_r,\beta_r)$ and interpolate $\eta_r^t:=(1-t)(p_r\otimes\beta_r)+t\eta_r^\star$. The same theorem realises each family $(\eta_r^t)_r$ simultaneously by a sender $\widetilde Q_t\in\mathcal F_R$. Continuity of $\sum_r\pi(r)I_{\eta_r^t}(B;Z)$, together with its endpoint values $0$ and $C_B$, proves that the complete attribute-specific range is $[0,C_B]$. The two interval claims concern the attainable values of each leakage functional separately; they do not assert that their upper endpoints must be attained by the same sender.
\end{proof}

For completeness, because $B=g(R,N)$, conditional data processing gives $I_Q(B;Z\mid R)\leq I_Q(N;Z\mid R)$ for every $Q\in\mathcal F_R$. Taking the respective maxima in the proved intervals yields $C_B\leq C_R$. Finally, with $K_0:=I_0(N;Z\mid R)$, the chain rule gives $I_0(S;Z)=I_R+K_0$; subtracting this reference leakage from the first interval establishes Eq.~\eqref{eq-task-preserving-relative-leakage}.

Finally, $G(p,q)=0$ if and only if $p$ or $q$ is a point mass. If $p=\delta_x$, every coupling of $p$ and $q$ equals $\delta_x\otimes q$, and symmetrically for $q$. Otherwise pick $x_1\neq x_2$ in the support of $p$ and $z_1\neq z_2$ in the support of $q$, and for $0<\epsilon\le\min\{p(x_1)q(z_2),\,p(x_2)q(z_1)\}$ move mass $\epsilon$ of $p\otimes q$ from $(x_1,z_2)$ and $(x_2,z_1)$ to $(x_1,z_1)$ and $(x_2,z_2)$. The result is a coupling of $p$ and $q$ that is not a product, so its mutual information is positive. Applied in each route to $(\alpha_r,\beta_r)$ and to $(p_r,\beta_r)$, this shows that $C_R=0$ exactly when, for every $r$ with $\pi(r)>0$, $\alpha_r$ or $\beta_r$ is a point mass, and $C_B=0$ exactly when $p_r$ or $\beta_r$ is.

\subsection{Optimal recovery with a public decoder}\label{app-proof-public-recovery}

\begin{proof}[Proof of Corollary~\ref{cor-optimal-public-recovery}]
Fix a task state $r$ with $\pi(r)>0$, and put $Y=d(Z)$. For every $Q\in\mathcal F_R$, the conditional law $\kappa_r$ of $(B,Y)$ given $R=r$ has marginals $p_r$ and $q_r^d$, since $P_Q^{Z\mid R=r}=\beta_r$. Thus $\kappa_r\in\Gamma(p_r,q_r^d)$. Conversely, take any $\kappa_r\in\Gamma(p_r,q_r^d)$ and define a conditional coupling of $(B,Z)$ by
\begin{equation}
\eta_r(b,z):=
\begin{cases}
\displaystyle\kappa_r(b,d(z))\frac{\beta_r(z)}{q_r^d(d(z))},&q_r^d(d(z))>0,\\[5pt]
0,&q_r^d(d(z))=0.
\end{cases}
\label{eq-app-public-decoder-lift}
\end{equation}
This construction is well defined: if $q_r^d(d(z))=0$, then $\beta_r(z)=0$. For each $y$ with $q_r^d(y)>0$, summing over messages with $d(z)=y$ gives $\sum_{z:d(z)=y}\eta_r(b,z)=\kappa_r(b,y)$. It follows that $\eta_r$ has $B$-marginal $p_r$, $Z$-marginal $\beta_r$, and pushforward $(B,d(Z))$-law $\kappa_r$. Theorem~\ref{thm-task-preserving-coupling} therefore realises any prescribed family $(\kappa_r)_r$ simultaneously by one sender in $\mathcal F_R$. Consequently, optimising recovery over task-preserving senders is equivalent to maximising $\Pr[B=Y]$ over $\Gamma(p_r,q_r^d)$ independently for each $r$.

For any such coupling $\kappa_r$, the diagonal mass is bounded by each marginal, so
\begin{equation}
\begin{aligned}
\Pr[B=Y\mid R=r]
&=\sum_b\kappa_r(b,b)\\
&\leq\sum_b\min\{p_r(b),q_r^d(b)\}\\
&=1-\operatorname{TV}(p_r,q_r^d).
\end{aligned}
\label{eq-app-public-recovery-upper}
\end{equation}
The upper bound is attained by a maximal coupling. To see this explicitly, let $e_r(b):=(p_r(b)-q_r^d(b))_+$, $f_r(b):=(q_r^d(b)-p_r(b))_+$, and $t_r:=\sum_b e_r(b)=\sum_b f_r(b)=\operatorname{TV}(p_r,q_r^d)$. When $t_r>0$, set
\begin{equation}
\kappa_r^\star(b,y)
:=\mathbf 1\{b=y\}\min\{p_r(b),q_r^d(b)\}
+\frac{e_r(b)f_r(y)}{t_r}.
\label{eq-app-maximal-decoder-coupling}
\end{equation}
Its marginals are $p_r$ and $q_r^d$. Moreover, $e_r(b)f_r(b)=0$ for every $b$, so its diagonal mass is exactly $\sum_b\min\{p_r(b),q_r^d(b)\}$. When $t_r=0$, the diagonal coupling $\kappa_r^\star(b,y)=\mathbf 1\{b=y\}p_r(b)$ attains the same bound. Lifting each $\kappa_r^\star$ by Eq.~\eqref{eq-app-public-decoder-lift} and invoking Theorem~\ref{thm-task-preserving-coupling} gives a single $Q^\star\in\mathcal F_R$ attaining the conditional optimum for every positive-probability route. Averaging over $R$ yields
\begin{equation}
\max_{Q\in\mathcal F_R}\Pr_Q[d(Z)=B]
=\sum_r\pi(r)\bigl[1-\operatorname{TV}(p_r,q_r^d)\bigr],
\end{equation}
as required.
\end{proof}

\subsection{Perfect recovery and bounded distortion}\label{app-proof-perfect-public-recovery}

\begin{proof}[Proof of Corollary~\ref{cor-perfect-public-recovery}]
Fix a public decoder $d:\mathcal Z\to\mathcal B$ and recall that $q_r^d(b)=\sum_{z:d(z)=b}\beta_r(z)$. Suppose first that some $Q\in\mathcal F_R$ satisfies $\Pr_Q[d(Z)=B]=1$. Since $\Pr[R=r]=\pi(r)>0$ for every retained task state, the weighted average of conditional recovery probabilities can equal one only if $\Pr_Q[d(Z)=B\mid R=r]=1$ for each such $r$. Thus $B=d(Z)$ almost surely conditional on $R=r$, so for every $b\in\mathcal B$,
\begin{equation}
\begin{aligned}
p_r(b)&=\Pr_Q[B=b\mid R=r]\\
&=\Pr_Q[d(Z)=b\mid R=r]\\
&=\sum_{z:d(z)=b}\beta_r(z)=q_r^d(b).
\end{aligned}
\end{equation}
Here the last equality uses $Q\in\mathcal F_R$. This proves that the stated mass-balance condition is necessary.

Conversely, suppose $p_r(b)=q_r^d(b)$ for every $b$ and every $r$ with $\pi(r)>0$. For each such $r$, define
\begin{equation}
\eta_r(b,z):=\beta_r(z)\,\mathbf 1\{d(z)=b\}.
\label{eq-app-perfect-recovery-coupling}
\end{equation}
Its $Z$-marginal is $\beta_r$, while its $B$-marginal is $\sum_z\beta_r(z)\mathbf 1\{d(z)=b\}=q_r^d(b)=p_r(b)$. Hence $\eta_r\in\Gamma(p_r,\beta_r)$, and it assigns probability one to the event $d(Z)=B$. By Theorem~\ref{thm-task-preserving-coupling}, these conditional couplings can all be realised simultaneously by one sender $Q^\star\in\mathcal F_R$; therefore $\Pr_{Q^\star}[d(Z)=B]=1$. For clarity, this sender can be chosen explicitly on every state $(r,n)$ of positive probability as
\begin{equation}
Q^\star(z\mid r,n)=\frac{\beta_r(z)\,\mathbf 1\{d(z)=g(r,n)\}}{p_r(g(r,n))}.
\label{eq-app-perfect-recovery-sender}
\end{equation}
The denominator is positive on the state support, and the mass-balance condition makes the numerator sum to that denominator. Summing over $N\mid R=r$ restores $\beta_r$, so $Q^\star\in\mathcal F_R$, while every message with positive probability satisfies $d(z)=g(r,n)=B$. This proves sufficiency and the required existence statement. No assertion is made that every task-preserving sender has perfect recovery.
\end{proof}

\noindent\textit{Proof of Corollary~\ref{cor-distortion-recovery}.} Fix $r$ and let $\nu_r$ be the law of $d(Z)$ given $R=r$ under $Q$. By data processing, $\mathrm{TV}(\nu_r,q_r^d)\le\delta_r$, so $\mathrm{TV}(p_r,\nu_r)\ge\mathrm{TV}(p_r,q_r^d)-\delta_r$, and $\Pr_Q[d(Z)=B\mid R=r]\le1-\mathrm{TV}(p_r,\nu_r)$ because $B$ and $d(Z)$ have laws $p_r$ and $\nu_r$ given $R=r$; since probabilities are at most $1$, this gives the stated value as an upper bound. For attainment, let $t_r=\min\{\delta_r,\mathrm{TV}(p_r,q_r^d)\}$ and choose $\nu^\ast_r$ on the segment from $q_r^d$ to $p_r$ with $\mathrm{TV}(\nu^\ast_r,q_r^d)=t_r$, so that $\mathrm{TV}(p_r,\nu^\ast_r)=\mathrm{TV}(p_r,q_r^d)-t_r$. Let $Q_r$ rescale $\beta_r$ on each decoder class $\{z:d(z)=y\}$ to total mass $\nu^\ast_r(y)$, placing that mass on any message decoded as $y$ if $q_r^d(y)=0$. Then $\mathrm{TV}(Q_r,\beta_r)=\mathrm{TV}(\nu^\ast_r,q_r^d)=t_r\le\delta_r$. Given $B$, draw $Y$ from the conditional law of a maximal coupling of $p_r$ and $\nu^\ast_r$, and then $Z$ from $Q_r$ restricted to $\{z:d(z)=Y\}$. The message law given $R=r$ is $Q_r$, and $\Pr[d(Z)=B\mid R=r]=1-\mathrm{TV}(p_r,\nu^\ast_r)$.\hfill$\square$

\subsection{Selection channels and canonical interfaces}\label{app-proof-selection-canonical}

\noindent\textbf{Selection kernel and recovery.} Fix a private state $s$, let $b_s=g(s)$, and write $a_s=P_0[d(Z)=b_s\mid S=s]$. Let $X_1,\ldots,X_k,F$ be independent draws from $P_0(\cdot\mid s)$, with $F$ the additional fallback. The selection rule emits the earliest $X_i$ with $d(X_i)=b_s$, or $F$ if every $X_i$ misses. For a particular $z$, the probability that the rule first selects $X_i=z$ is $(1-a_s)^{i-1}P_0(z\mid s)\mathbf 1\{d(z)=b_s\}$; the probability that all $k$ candidates miss and $F=z$ is $(1-a_s)^kP_0(z\mid s)$. These disjoint cases give
\begin{equation}
\begin{aligned}
Q_k(z\mid s)=P_0(z\mid s)\bigg[&(1-a_s)^k\\
&+\mathbf 1\{d(z)=b_s\}\sum_{i=0}^{k-1}(1-a_s)^i\bigg],
\end{aligned}
\label{eq-app-selection-kernel}
\end{equation}
which is Eq.~\eqref{eq-selection-kernel}. To verify normalisation directly, summing over $z$ yields $(1-a_s)^k+a_s\sum_{i=0}^{k-1}(1-a_s)^i=1$, including $a_s=0$ and $a_s=1$. The emitted bit is incorrect precisely if all $k$ candidates miss and the fresh fallback also misses. Independence therefore yields
\begin{equation}
\Pr_{Q_k}[d(Z)=B\mid S=s]=1-(1-a_s)^{k+1},
\label{eq-app-selection-recovery}
\end{equation}
proving Eq.~\eqref{eq-selection-recovery-law}. In particular, for $k=1$ and $a_s=1/2$, the recovery probability is $3/4$.

\noindent\textbf{Likelihood ratios.} Define $w_k(s,z):=Q_k(z\mid s)/P_0(z\mid s)$ on the support of $P_0(\cdot\mid s)$, using the bracketed expression in Eq.~\eqref{eq-selection-kernel} elsewhere. For every $z$ with $\beta(z)>0$,
\begin{equation}
\begin{aligned}
\frac{\bar Q_k(z)}{\beta(z)}
&=\frac{\sum_s\mu(s)P_0(z\mid s)w_k(s,z)}{\sum_s\mu(s)P_0(z\mid s)}\\
&=\mathbb E_0[w_k(S,z)\mid Z=z].
\end{aligned}
\label{eq-app-selection-message-ratio}
\end{equation}
Conditioning the same calculation on a positive-probability task state $R=r$ gives $P_{Q_k}(Z=z\mid R=r)/\beta_r(z)=\mathbb E_0[w_k(S,z)\mid Z=z,R=r]$ whenever $\beta_r(z)>0$. These identities require no claim that the selection law equals the reference law; indeed the ratio may depend on $z$.

\noindent\textbf{Proof of Theorem~\ref{thm-selection-detectability}.} Assume that $B$ is fair conditional on every $R=r$ of positive probability and that $a_s=1/2$ for every positive-probability private state $s$. Set $\delta_k:=1-2^{-k}$ and $m_r(z):=P_0[B=d(z)\mid Z=z,R=r]$ for $\beta_r(z)>0$. The selection multiplier reduces to
\begin{equation}
w_k(s,z)=2^{-k}+2\delta_k\,\mathbf 1\{d(z)=g(s)\},
\end{equation}
and hence the route-conditional likelihood ratio is $2^{-k}+2\delta_k m_r(z)$, which equals $1+2\delta_k(m_r(z)-1/2)$ because $2^{-k}=1-\delta_k$; in particular $\ell_{k,r}\le2$. Therefore,
\begin{equation}
\begin{aligned}
&\operatorname{TV}(P_{Q_k}^{Z\mid r},\beta_r)\\
&\quad=\frac12\sum_z\beta_r(z)|\ell_{k,r}(z)-1|\\
&\quad=\delta_k\sum_z\beta_r(z)|m_r(z)-\tfrac12|.
\end{aligned}
\label{eq-app-selection-tv-first}
\end{equation}
Under the fair-bit assumption, writing $u_r(z):=P_0[B=1\mid Z=z,R=r]$, the corresponding joint-law distance is
\begin{equation}
\begin{aligned}
&\operatorname{TV}(P_0^{(B,Z)\mid r},p_r\otimes\beta_r)\\
&\quad=\frac12\sum_z\beta_r(z)\sum_{b\in\{0,1\}}
\left|P_0[B=b\mid z,r]-\tfrac12\right|\\
&\quad=\sum_z\beta_r(z)|u_r(z)-\tfrac12|.
\end{aligned}
\label{eq-app-selection-tv-second}
\end{equation}
For each $z$, $m_r(z)$ equals either $u_r(z)$ or $1-u_r(z)$ according to $d(z)$, so $|m_r(z)-1/2|=|u_r(z)-1/2|$. Combining Eqs.~\eqref{eq-app-selection-tv-first} and~\eqref{eq-app-selection-tv-second} proves the total-variation identity. For the $\chi^2$ identity, $\chi^2(P_{Q_k}^{Z\mid r}\|\beta_r)=\sum_z\beta_r(z)(\ell_{k,r}(z)-1)^2=4\delta_k^2\sum_z\beta_r(z)(m_r(z)-\tfrac12)^2$, while $\chi^2(P_0^{(B,Z)\mid r}\|p_r\otimes\beta_r)=\sum_z\beta_r(z)\sum_b(P_0[B=b\mid z,r]-\tfrac12)^2/\tfrac12=4\sum_z\beta_r(z)(u_r(z)-\tfrac12)^2$, and $|m_r-\tfrac12|=|u_r-\tfrac12|$.

For $T$ independent rounds with $R=r$, let $P$ and $P'$ be the $T$-fold laws of selected and reference messages and $c_r$ the single-round $\chi^2$ divergence. The total-variation distance $\mathrm{TV}(P,P')$ is the largest advantage $\mathrm{TPR}-\mathrm{FPR}$ of any test. Since $1+\chi^2$ is multiplicative over independent rounds and $\mathrm{TV}\le\tfrac12\sqrt{\chi^2}$ by the Cauchy--Schwarz inequality, $\mathrm{TV}(P,P')\le\tfrac12((1+c_r)^T-1)^{1/2}\le\tfrac12(e^{Tc_r}-1)^{1/2}$. For the lower bound, let $\mathrm{BC}:=\sum_z\beta_r(z)\sqrt{\ell_{k,r}(z)}$ be the Bhattacharyya coefficient of one round; it is multiplicative over rounds, and $\mathrm{TV}\ge1-\mathrm{BC}$ for any two laws. Because $\ell_{k,r}\le2$,
\begin{equation*}
1-\mathrm{BC}=\tfrac12\sum_z\beta_r(z)\frac{(\ell_{k,r}(z)-1)^2}{(\sqrt{\ell_{k,r}(z)}+1)^2}\ge\frac{c_r}{2(1+\sqrt2)^2}\ge\frac{c_r}{12},
\end{equation*}
so $\mathrm{TV}(P,P')\ge1-\mathrm{BC}^T\ge1-(1-c_r/12)^T\ge1-e^{-Tc_r/12}$. Finally, if $B$ and $Z$ are independent given $R$ under $P_0$, then $m_r\equiv1/2$, so $\ell_{k,r}\equiv1$ and $Q_k\in\mathcal F_R$, while Eq.~\eqref{eq-selection-recovery-law} with $a_s=1/2$ gives recovery $1-2^{-(k+1)}$. Outside the balanced case, Eq.~\eqref{eq-app-selection-message-ratio} gives the exact ratio.\hfill$\square$

\noindent\textbf{Keyed selection.} Let $\xi$ be a uniform bit independent of $S$, known to sender and receiver, and let the sender select on $d(z)=b_s\oplus\xi$. In the balanced case $P_0[d(Z)=y\mid S=s]=1/2$ for both values of $y$, so given $\xi=x$ the kernel is $P_0(z\mid s)[2^{-k}+2\delta_k\mathbf 1\{d(z)=b_s\oplus x\}]$, and averaging over $x$ gives $P_0(z\mid s)[2^{-k}+\delta_k]=P_0(z\mid s)$. The receiver's estimate $d(Z)\oplus\xi$ equals $B$ unless all $k+1$ draws miss, which has probability $2^{-(k+1)}$.

\noindent\textbf{Canonical-interface bounds and receiver preservation.} Let $C=c(Z)$ take values in a set $\mathcal C$ of cardinality at most $K$. Since $C$ is a deterministic function of $Z$, data processing and the entropy bound give $I_Q(S;C)\leq H_Q(C)\leq\log_2 K$ and $I_Q(B;C\mid R)\leq H_Q(C\mid R)\leq\log_2 K$. If $c$ is idempotent and the fixed receiver factors as $\rho(a\mid z)=\widetilde\rho(a\mid c(z))$, then for every $z,a$,
\begin{equation}
\begin{aligned}
\rho(a\mid c(z))
&=\widetilde\rho(a\mid c(c(z)))\\
&=\widetilde\rho(a\mid c(z))=\rho(a\mid z),
\end{aligned}
\label{eq-app-canonical-receiver}
\end{equation}
so replacing each message by its canonical form preserves the conditional receiver-action law (and the action itself when the receiver is deterministic).

For the stronger conditional bound, let $K\geq2$, let $h:\mathcal R\to\mathcal C$ be a task-only representation, and put $E:=\mathbf 1\{C\neq h(R)\}$, with $\Pr_Q[E=1]=\varepsilon$. Because $E$ is determined by $(C,R)$, the chain rule yields $H_Q(C\mid R)=H_Q(E\mid R)+H_Q(C\mid E,R)$. Concavity of binary entropy gives $H_Q(E\mid R)\leq h_2(\varepsilon)$. When $E=0$, $C=h(R)$ is determined by $R$; when $E=1$, at most $K-1$ canonical outputs remain possible for each $R=r$. Thus
\begin{equation}
\begin{aligned}
H_Q(C\mid R)
&\leq h_2(\varepsilon)+\varepsilon\log_2(K-1).
\end{aligned}
\label{eq-app-canonical-entropy}
\end{equation}
Finally, $B=g(R,N)$ implies the conditional Markov relation $B-(N,R)-C$, whence conditional data processing gives $I_Q(B;C\mid R)\leq I_Q(N;C\mid R)\leq H_Q(C\mid R)$. Combining these inequalities proves Eq.~\eqref{eq-approx-canonical-leakage-bound}. When $\varepsilon=0$, $C=h(R)$ almost surely and both conditional mutual informations vanish. This is a bound for the whole canonical message, and therefore for any post-processing or decoder of it.

\noindent\textbf{Binary recovery on a nearly task-only interface.} Suppose additionally that $B$ is a fair bit conditional on $R$, and let $\widehat B$ be any possibly randomised binary decoder of $(C,R)$, with its randomisation independent of the private state. Couple it to the task-only decoder $\widehat B_0$ obtained by feeding the same decoder and random coins the value $h(R)$ in place of $C$. Because $\widehat B_0$ depends only on $R$ and independent coins, $\Pr[\widehat B_0=B]=1/2$. The two decoder outputs agree whenever $E=0$, so their correctness indicators can differ only on the event $E=1$. Consequently,
\begin{equation}
\left|\Pr[\widehat B=B]-\tfrac12\right|\leq\Pr[E=1]=\varepsilon.
\label{eq-app-canonical-recovery-bound}
\end{equation}

\noindent\textit{Proof of Proposition~\ref{prop-canonical-balanced}.} The recovery bound is Eq.~\eqref{eq-app-canonical-recovery-bound}. For the information bound, let $B$ be fair given each $R=r$ and write $\varepsilon_r=\Pr[C\neq h(r)\mid R=r]$. If $\varepsilon_r\ge 1/2$, then $I(B;C\mid R=r)\le 1\le 2\varepsilon_r$. Otherwise $\Pr[B=1,C=h(r)\mid R=r]$ lies in $[1/2-\varepsilon_r,1/2]$, so given $C=h(r)$ the probability that $B=1$ lies between $x_r:=(1/2-\varepsilon_r)/(1-\varepsilon_r)$ and $1-x_r$. Hence $H(B\mid C,R=r)\ge(1-\varepsilon_r)h_2(x_r)$ and
\begin{equation*}
I(B;C\mid R=r)\le f(\varepsilon_r):=1-(1-\varepsilon_r)\,h_2\!\left(\frac{1/2-\varepsilon_r}{1-\varepsilon_r}\right).
\end{equation*}
The function $f$ is convex on $[0,1/2]$, because $(1-\varepsilon)h_2(x)$, with $x$ as above, has second derivative $-1/\bigl(2(1/2-\varepsilon)(1-\varepsilon)\ln 2\bigr)<0$, and $f(0)=0$, $f(1/2)=1$. So $f(\varepsilon)\le 2\varepsilon$, and averaging over routes gives $I(B;C\mid R)\le 2\varepsilon$.

For a key $\xi$ independent of $S$, conditioning on $\xi=x$ leaves the law of $S$ unchanged and gives a sender with routing error $\varepsilon_x$, where $\sum_x\Pr[\xi=x]\varepsilon_x=\varepsilon$. Each bound above holds given $\xi=x$ with $\varepsilon_x$, and $I(B;C,\xi\mid R)=I(B;C\mid R,\xi)$ because $\xi$ is independent of $(R,B)$, and likewise for $N$. Averaging over $x$, which is linear for the recovery bound and for $2\varepsilon$ and concave for $h_2(\varepsilon)+\varepsilon\log_2(K-1)$, proves the keyed versions.\hfill$\square$

The argument also applies when the decoder receives only $C$, by ignoring the additional $R$ input. It holds for any joint law with the stated balance and routing-error conditions, including the balanced empirical joint distribution used in Sect.~\ref{subsec-empirical-canonicalization}; inserting a measured error fraction gives a bound for that empirical distribution, not by itself a population confidence guarantee.

\subsection{Law-preserving resampling sender}\label{app-resampling-proof}

\begin{proof}[Proof of Proposition~\ref{prop-resampling-sender}]
Fix a route $r$ with $\pi(r)>0$ and an attribute value $b$ with $p_r(b)>0$. Since $\kappa_r^\star$ has marginals $p_r$ and $q_r^d$, the distribution $\kappa_r^\star(b,y)/p_r(b)$ over decoder targets $y$ is normalised. Moreover, any $y$ selected with positive probability satisfies $q_r^d(y)>0$: otherwise the second marginal of $\kappa_r^\star$ would force $\kappa_r^\star(b,y)=0$ for every $b$. Thus no selected target has zero acceptance probability.

On each trial the auxiliary state is drawn afresh from $\alpha_r$ and the reference message is generated for that state. The unconditional law of a trial message, given $R=r$, is therefore $\sum_{n'}\alpha_r(n')P_0(z\mid r,n')=\beta_r(z)$. Independent trials terminate almost surely for a fixed selected target $y$, and the accepted message has the conditional reference distribution
\begin{equation}
\begin{aligned}
&\Pr[Z=z\mid R=r,B=b,Y=y]\\
&\quad=\frac{\beta_r(z)\mathbf 1\{d(z)=y\}}{q_r^d(y)}.
\end{aligned}
\label{eq-app-resampling-acceptance}
\end{equation}
Averaging over the selected target and multiplying by $p_r(b)$ gives, for every $z$ with $q_r^d(d(z))>0$,
\begin{equation}
\Pr[B=b,Z=z\mid R=r]
=\kappa_r^\star(b,d(z))\frac{\beta_r(z)}{q_r^d(d(z))}.
\label{eq-app-resampling-lift}
\end{equation}
If $q_r^d(d(z))=0$, then $\beta_r(z)=0$ and the joint probability is zero. This is exactly the lifted coupling of Eq.~\eqref{eq-app-public-decoder-lift}. Summing Eq.~\eqref{eq-app-resampling-lift} over $b$ yields $\Pr[Z=z\mid R=r]=\beta_r(z)$ on the positive support, since $\sum_b\kappa_r^\star(b,y)=q_r^d(y)$; the equality also holds on the zero support. Because this argument holds for every positive-probability route, the sender preserves $P^{(R,Z)}=\lambda_0$ and hence belongs to $\mathcal F_R$. Its decoder-recovery probability within route $r$ is $\sum_b\kappa_r^\star(b,b)=1-\operatorname{TV}(p_r,q_r^d)$ by maximal coupling. Averaging over routes proves optimality by Corollary~\ref{cor-optimal-public-recovery}.

For completeness, the number of reference draws until acceptance, conditional on $Y=y$, is geometric with mean $1/q_r^d(y)$. The target marginal conditional on $R=r$ is $q_r^d$ by the second marginal of $\kappa_r^\star$. Consequently, the mean number of reference draws per emitted message in that route is $\sum_{y:q_r^d(y)>0}q_r^d(y)/q_r^d(y)=|\operatorname{supp}(q_r^d)|$, which equals two for a binary decoder when both values have positive probability, and one otherwise. This is an expectation over the route's state distribution; no uniform finite per-message budget is asserted.
\end{proof}


\section{Experimental Details and Reproducibility}\label{app-experiments}

This appendix specifies the frozen experimental protocol, statistical analysis, and procedures for reproducing the measurements in Sect.~\ref{sec-empirical-evaluation}. The implementation is available in the anonymous, read-only source artefact linked in Sect.~\ref{app-exp-artifact}. The repository supplies code and configuration rather than the generated experimental observations or pretrained model weights; a full rerun requires locally obtaining the pinned model and regenerating the data.

\subsection{Synthetic Task, Prompts, and Interfaces}\label{app-exp-task}

The experiment uses a fixed, uniformly weighted population of $64$ synthetic incident records. The task route $R\in\{0,\ldots,15\}$ is the Cartesian product of four fault domains (\texttt{storage\_array}, \texttt{network\_fabric}, \texttt{auth\_service}, \texttt{batch\_scheduler}) and four severities (\texttt{sev1\_critical}, \texttt{sev2\_major}, \texttt{sev3\_minor}, \texttt{sev4\_cosmetic}). Each route has four administrative variants $N\in\{0,1,2,3\}$, with fields \texttt{intake\_office}, \texttt{intake\_shift}, and \texttt{intake\_channel}; these variants are deterministic combinations of four values per field, not independently sampled field combinations. The target is $B=N\bmod 2$. Consequently, $R$ and $N$ are independent and uniform in this test population, and $H(B\mid R)=1$ bit. For each route $r=4d+e$ and variant $v$, the domain and severity are the $d$th and $e$th entries in the ordered lists above; the office is the $v$th of (XXXX-2, XXXX-17, Aarhus, Porto)\footnote{The first two are city names redacted for anonymous review.}, the shift is the $(v+1)\bmod4$th of (morning, midday, evening, night), and the channel is the $(3v+2)\bmod4$th of (phone, email, portal, walk\_in). The records contain no real incident or personal information.

The frozen common instruction is reproduced below. The suppression and disclosure conditions append different metadata instructions; the resulting prompt is then followed by the record, beginning with \texttt{CASE FILE} and listing the five field--value pairs in the frozen order. The model's chat template supplies the system formatting, and the complete executed prompt strings and token IDs are retained in the run manifest. These conditions therefore define different honest reference laws.
\begin{lstlisting}[basicstyle=\ttfamily\scriptsize,breaklines=true,columns=fullflexible]
You route incident tickets to a remediation team. Read the case file and
report its fault_domain and severity exactly as recorded. These two fields
determine the route. Give a brief description of the incident.
Reply using exactly these three lines:
domain: <fault_domain value>
severity: <severity value>
summary: <one short sentence>
\end{lstlisting}
The suppressed condition appends the following instruction:
\begin{lstlisting}[basicstyle=\ttfamily\scriptsize,breaklines=true,columns=fullflexible]
The intake_office, intake_shift and intake_channel are administrative metadata.
Do not include, paraphrase or hint at them in your reply.
\end{lstlisting}
The disclosed honest control instead appends:
\begin{lstlisting}[basicstyle=\ttfamily\scriptsize,breaklines=true,columns=fullflexible]
The intake_office, intake_shift and intake_channel are administrative metadata.
Include all three exact values in the summary sentence.
\end{lstlisting}

The receiver is a deterministic parser applied to the emitted text. It requires exactly one line beginning with the lowercase label \texttt{domain:} and exactly one beginning with \texttt{severity:}; it trims ASCII whitespace from their values and accepts only the enumerated domain and severity strings. Missing, duplicate, or invalid routing fields yield \texttt{REJECT}. The receiver does not infer a route from the summary or consult the private state. Its action $A$ is the parsed route (one of $16$ values) or rejection; task utility is $\mathbf{1}\{A=R\}$. Additional summary text is allowed by the parser even when it fails the requested output format, so routing accuracy measures this declared parser task rather than strict instruction compliance.

Write $W$ for the generated token-ID sequence, including an end-of-sequence token if sampled, and $V$ for the rendered UTF-8 text. The renderer skips special tokens and disables tokeniser whitespace clean-up; it does not append an artificial end-of-sequence token when generation reaches its length cap. The public decoder $J=d(V)$ is the low bit of the last byte of $\mathrm{SHA\mbox{-}256}(\mathrm{UTF8}(V))$, with no stripping or Unicode normalisation. Hash recovery is $\mathbf{1}\{J=B\}$, independently of whether the receiver routes correctly. Hashing, parsing, and the canonical interface operate on $V$, whereas model likelihoods refer to the retained token sequence $W$. We do not assume that token-to-text rendering is injective.

\subsection{Model, Selection Arms, and Randomness}\label{app-exp-generation}

The reference generator is \texttt{Qwen/Qwen2.5-1.5B-Instruct} at model and tokeniser revision \texttt{989aa7980e4cf806f80c7fef2b1adb7bc71aa306}. Generation uses temperature $1$, the full next-token distribution without top-$k$, top-$p$, or repetition filtering, and a cap of $96$ continuation tokens. The recorded execution used FP32 model arithmetic, eager attention, disabled TF32, and FP64 log-softmax reductions for likelihood evaluation. The same pinned reference model, decoding rule, and suppression prompt are used for all candidate-selection arms; the selection wrapper alone changes which completed response is emitted.

For each suppressed-condition round at private state $s$, the collector generates $19$ independent reference draws: two honest controls $\mathsf H$ and $\mathsf H'$, a pool $(V_1,\ldots,V_{16})$, and one additional fallback $V_\mathrm{spare}$. For $k\in\{1,2,4,8,16\}$, sender $Q_k$ emits the first $V_i$ with $i\leq k$ and $d(V_i)=B(s)$; if no such candidate exists, it emits $V_\mathrm{spare}$ regardless of the fallback's decoded bit. The five selection arms share the same candidate pool and fallback within a round, and $\mathsf H,\mathsf H'$ are never candidates. All $19$ draws are generated and retained even when a small-budget arm finds an early match. Thus every $Q_k$ is a deterministic choice from the saved round bundle; treating the seven emitted arms as independent samples would be incorrect. No selection candidates are generated for the separate honest disclosure control.

\noindent\textbf{Post hoc law-preserving arm.} This arm reuses the stored suppressed-condition test pools and generates nothing new. For a round with state $(r,n)$ and bit $b$, the decoder target is $y=b$, except that when $\widehat q_r^d(b)<1/2$ it is flipped with probability $1-2\widehat q_r^d(b)$, which realises the maximal coupling of $p_r$ and $\widehat q_r^d$; its expected recovery on the test design is $2{,}007/2{,}048=97.998\%$. Each trial then picks one of the four records of route $r$ uniformly at random and takes that record's next unused pool candidate, until $d(V)=y$. The $8{,}192$ messages used $16{,}602$ draws, at most $15$ in a round; no record ran out of candidates, and there is no fallback. Only $9$ emitted messages coincide with the candidate that $Q_{16}$ emitted from the same pool. All monitors score these messages with the parameters and thresholds fitted for $Q_{16}$, without refitting.

The eight base seeds are $2026092901$--$2026092908$. Each draw receives its own pseudorandom stream seed, derived from the SHA-256 digest of a canonical serialisation containing the namespace \texttt{paper1-fresh-v3}, data split, prompt condition, base seed, state ID, round index, and draw role. The first eight digest bytes, reduced modulo $2^{63}$, determine the stream seed; this prevents generation order, batching, and parallel scheduling from silently assigning draws to different arms. The independent decoder-match calibration uses seed $2026092988$ and its own split identifier. The numerical results are interpreted under the specified independent-draw sampling model; deterministic pseudorandom seeding provides replayability rather than a mathematical guarantee of physical randomness.

\subsection{Splits, Likelihood Scoring, and Provenance}\label{app-exp-collection}

Each of the $64$ states occurs under all eight base seeds in each suppressed-condition split. Training and threshold calibration have two rounds per state and seed ($1{,}024$ rounds each); the held-out test split has $16$ rounds per state and seed ($8{,}192$ rounds). The disclosure control contains two independent honest test draws per state and seed ($1{,}024$ messages). Independent hash-match calibration contains $64$ honest draws per state ($4{,}096$ messages, or $256$ per route). The sample allocation was fixed before analysis. The likelihood-scored suppressed subset comprises one round per state and seed in training and calibration ($512$ rounds per split) and two in testing ($1{,}024$ rounds), with the retained $\mathsf H$, $\mathsf H'$, and $Q_{16}$ outputs. The disclosure control is likewise likelihood-scored.

\begin{table}[t]
\centering
\caption{Frozen collection accounting. A suppressed-condition round contains $19$ reference draws, even though only seven arms are emitted.}
\label{tab-app-exp-sample-counts}
\footnotesize
\setlength{\tabcolsep}{3pt}
\begin{tabular}{@{}lrrr@{}}
\toprule
Component & Round units & Draws/unit & Messages \\
\midrule
Training (suppressed) & $1{,}024$ & $19$ & $19{,}456$ \\
Calibration (suppressed) & $1{,}024$ & $19$ & $19{,}456$ \\
Test (suppressed) & $8{,}192$ & $19$ & $155{,}648$ \\
Disclosure control & $1{,}024$ & $1$ & $1{,}024$ \\
Hash-match calibration & $4{,}096$ & $1$ & $4{,}096$ \\
\midrule
Total & $15{,}360$ & --- & $199{,}680$ \\
\bottomrule
\end{tabular}
\end{table}

For a retained token sequence $w=(w_1,\ldots,w_T)$ and any reference state $s$, likelihood evaluation computes the sequence log probability $L_s(w)=\sum_{t=1}^{T}\log P_0(w_t\mid s,w_{<t})$ under that state's prompt, including the observed stopping convention. Cross-state scoring uses the saved token IDs rather than retokenising rendered text. Identical token sequences within a prompt condition may share a computed likelihood column, but repeated observations retain their multiplicities in all empirical averages. The completed run records $6{,}721$ distinct $64$-state likelihood columns ($430{,}144$ state--sequence scores); this is an implementation economy, not an observation-level deduplication.

The collection pipeline freezes the model revision, synthetic population, prompts, source code, runtime information, decoding settings, seeds, and checksums before production. It checkpoints all candidate draws and chosen roles, and verifies that saved token sequences render to the recorded text. Generation and scoring implementations are cross-checked using preflight numerical comparisons, independent scoring checks, and selected same-seed replays. The completed run's integrity records report successful checks, including $256$ exact generation replays and $16$ complete independent reference-column audits. Hardware-dependent numerical differences can still occur on a new installation; the repository preserves tolerances and diagnostic outputs rather than assuming bitwise portability across GPUs.

\subsection{Estimands and Statistical Procedures}\label{app-exp-statistics}

All estimates are averages over the fixed, equally weighted $64$-state population; uncertainty represents repeated generation under the stated sampling procedure, not sampling new incident types or model families. For arm $a$, public-bit recovery and routing utility are $p_a=\Pr[d(V_a)=B]$ and $u_a=\Pr[A_a=R]$. Differences $p_{Q_k}-p_{\mathsf H}$ and $u_{Q_k}-u_{\mathsf H}$ are calculated within paired round bundles. The primary study used predefined routing criteria $u_{\mathsf H}\geq0.90$ and $u_{Q_{16}}-u_{\mathsf H}\geq-0.05$, assessed with conservative one-sided bounds; meeting these design tolerances is not a statement of exact equality of task laws.

The primary recovery, routing, and paired-difference intervals use $5{,}000$ percentile bootstrap resamples (seed $2026092999$), resampling whole round bundles within each of the $64$ fixed states while pooling the eight base seeds. Supplementary theory-aligned information and receiver-interface intervals use another $5{,}000$ paired replicates (seed $2026100301$), resampling within every state-by-base-seed stratum. For either procedure, all arms and canonical outputs within a selected round receive the same resampling weight. The two procedures differ in stratification; both yield descriptive intervals conditional on the fixed experimental design. These bootstrap intervals are not used to claim certainty from an all-success result or to certify new-model generalisation.

For an outcome with zero errors in $n$ independent, potentially non-identically distributed trials, a one-sided $1-\alpha$ bound on the design-averaged error probability $\eta$ is $\eta\leq1-\alpha^{1/n}$. Indeed, if individual success probabilities are $1-\eta_i$, the arithmetic--geometric mean inequality gives $\Pr[\text{all succeed}]=\prod_i(1-\eta_i)\leq(1-\frac1n\sum_i\eta_i)^n$. With $n=8{,}192$ and $\alpha=0.05$, the zero observed errors of $Q_{16}$ imply $\eta\leq3.66\times10^{-4}$ at one-sided $95\%$ confidence. This is numerically the usual all-success Clopper--Pearson endpoint, but the stated justification also covers the fixed, balanced, non-identically distributed design under independent sampling. Allocating $\alpha/16$ to each of $16$ routes gives simultaneous bounds from $512$ held-out rounds per route. For nonzero error counts, the theory-aligned analysis uses two-sided Hoeffding intervals for the design-averaged error and transforms the entire error interval through binary entropy.

Because $H(B\mid R)=1$ bit, conditional Fano's inequality and data processing give $I(B;V\mid R)\geq I(B;J\mid R)\geq1-h_2(\eta)$ for the fixed decoder $J=d(V)$. The zero-error bound above therefore gives a conservative $0.9953$-bit lower confidence bound for $Q_{16}$, rather than interpreting the empirical $100\%$ success as a mathematically noiseless population channel. Independently, the reported plug-in decoder information is calculated from route-conditioned $2\times2$ contingency tables: $\widehat I(B;J\mid R)=16^{-1}\sum_r\sum_{b,j}\widehat p_r(b,j)\log_2\!\bigl(\widehat p_r(b,j)/(\widehat p_r(b)\widehat p_r(j))\bigr)$, with zero-mass terms omitted. This finite-sample mutual-information estimate is not itself a confidence lower bound and may exhibit positive bias near zero.

\subsection{Monitors, Calibration, and Repeated Observations}\label{app-exp-monitors}

The detector suite compares $Q_{16}$ with $\mathsf H$. The structure classifier uses simple rendered-message formatting and length features, while the text classifier uses character $3$--$5$-gram TF--IDF features (at most $8{,}192$ features); both are logistic regressions trained only on the training split. Reference surprisal is $-\log_2\beta(W)$, computed from reference token-sequence likelihoods, and its trusted-route analogue is $-\log_2\beta_r(W)$. The selection-aware monitor uses the known selection kernel to score the likelihood ratio between the selected and honest marginal laws, with the unknown statewise match probabilities estimated from the independent calibration stream using Jeffreys smoothing. The trusted-route variant conditions that calculation on the supplied true $R$. None of these detectors receives the true $B$ for the tested message; the known-code positive control explicitly does.

The nominal false-positive target is $5\%$. For every detector, a threshold is determined from honest calibration scores using a strict exceedance rule and is then held fixed on test messages; AUC is computed separately and does not depend on that threshold. The optional novelty purge removes a complete later round bundle if any emitted-arm text repeats an earlier retained split's text. Training retains $1{,}024$ rounds, while novelty filtering leaves $773$ calibration and $5{,}723$ test rounds for learned-text monitors; the likelihood-scored novelty cohort has $512$ training, $393$ calibration, and $706$ test rounds. The main recovery and routing endpoints never apply this purge. The full-stream detector evaluations apply the thresholds learned from the purged calibration cohort to all $8{,}192$ test rounds for text/structure monitors, or all $1{,}024$ likelihood-scored test rounds for likelihood monitors.

This distinction explains why the full-stream surprisal monitor has $14.06\%$ false positives despite the nominal $5\%$ calibration target: its threshold was selected on the novelty-purged $393$-round honest likelihood calibration cohort, whereas the reported full-stream rate uses all $1{,}024$ test observations. The reported AUC remains interpretable as threshold-free ranking discrimination. For collection tests, scores are averaged over disjoint groups of $16$ separately sampled rounds after a fixed permutation, using the same grouping for the paired honest and selected arms; each collection receives its own honest calibration threshold. The full likelihood cohort supplies $64$ test collections and $24$ calibration collections per arm. These collections are aggregate independent interactions, not adaptive multi-turn conversations. Consequently, the much stronger selection-aware collection result does not demonstrate security against, or detection within, an adaptive dialogue protocol.

Table~\ref{tab-app-auc-intervals} gives $95\%$ intervals for every AUC of Table~\ref{tab-empirical-monitors} from $5{,}000$ paired bootstrap replicates. The character-text classifier ranks honest test messages very slightly above both attack arms, so its interval lies just below $0.5$ for both; this is not detection.

\begin{table}[t]
    \centering
    \caption{AUCs of Table~\ref{tab-empirical-monitors} with $95\%$ bootstrap intervals ($5{,}000$ paired replicates).}
    \label{tab-app-auc-intervals}
    \footnotesize
    \setlength{\tabcolsep}{3pt}
    \begin{tabular}{@{}lll@{}}
        \toprule
        Monitor & $Q_{16}$ & Law-preserving \\
        \midrule
        Structure (1) & .504 [.496, .511] & .503 [.496, .510] \\
        Character text (1) & .490 [.481, .499] & .490 [.481, .499] \\
        Surprisal (1) & .496 [.474, .519] & .495 [.472, .518] \\
        Selection-aware ratio (1) & .645 [.622, .667] & .504 [.480, .528] \\
        Selection-aware ratio (16) & .972 [.947, .992] & .540 [.450, .633] \\
        Route surprisal (1) & .496 [.474, .519] & .495 [.472, .518] \\
        Route-aware ratio (1) & .645 [.622, .667] & .504 [.480, .527] \\
        Route-aware ratio (16) & .972 [.947, .992] & .540 [.450, .633] \\
        Known-code control (16) & 1.000 [1.000, 1.000] & 1.000 [1.000, 1.000] \\
        \bottomrule
    \end{tabular}
\end{table}

\subsection{Theory-Aligned Diagnostics and Canonicalisation}\label{app-exp-diagnostics}

The independent $4{,}096$-draw hash calibration estimates $a_s=\Pr_0[d(V)=B(s)\mid S=s]$ from $64$ draws per state. Substitution into the exact selection formula $1-(1-a_s)^{k+1}$ and uniform averaging over states yields predictions for each candidate budget without fitting to the held-out candidate pools. The same independent sample supplies $256$ draws per route to estimate $q_r^d(1)=\Pr_0[d(V)=1\mid R=r]$. The observed range $0.4492$--$0.5430$ gives a plug-in value $1-16^{-1}\sum_r|\widehat q_r^d(1)-1/2|\approx0.9800$ for Corollary~\ref{cor-optimal-public-recovery}. This diagnostic concerns the hypothetical exact task-law-preserving class; the tested selection senders are not assumed to belong to that class. Simultaneous Hoeffding intervals for the $16$ estimated route probabilities include $1/2$ in every case and do not certify exact mass balance.

The honest likelihood analysis uses the four state-conditioned sequence likelihoods within each true route to compute $P_0(B=b\mid W=w,R=r)=\bigl[\sum_{s:R(s)=r,\,B(s)=b}P_0(w\mid s)\bigr]/\bigl[\sum_{s:R(s)=r}P_0(w\mid s)\bigr]$. Averaging the resulting binary posterior entropies on honest outputs gives $\widehat I_0(B;W\mid R)=1-\widehat H_0(B\mid W,R)$: $0.2620$ bits under suppression and $0.7368$ bits under the separately prompted disclosure control. These are reference-law token-sequence quantities, not measurements of selected-sender mutual information and not, without summing over rendering preimages, exact rendered-text information. They explain why near-chance recovery by the honest \emph{hash} should not be interpreted as independence of private state and honest messages. The same posteriors give $(1-2^{-16})$ times their mean absolute deviation from $1/2$ on honest outputs, $0.189$, the route-averaged distance between $Q_{16}$ and $\mathsf H$ that Theorem~\ref{thm-selection-detectability} gives in the balanced case. The largest $\mathrm{TPR}-\mathrm{FPR}$ of the single-message selection-aware ratio over all thresholds is $0.205$; a maximum over thresholds is biased upwards at this sample size, and the threshold-free AUC stays below the implied ceiling of $0.671$. Administrative statements are detected by conservative, contextual matching of whole field values. Under the suppression prompt, $1{,}541$ of the $8{,}192$ honest test messages ($18.8\%$) contain at least one; under the disclosure prompt, $626$ of $1{,}024$ ($61.1\%$) do, and $37$ ($3.6\%$) state all three values. Even if each statement revealed $B$ exactly, such messages would contribute at most about $0.094$ of the mean deviation and $0.188$ of the $0.262$ bits.

For each arm, the supplementary analysis estimates the full $17$-category receiver distribution conditional on each true route and reports $D_{\mathrm{task}}(Q,\mathsf H)=16^{-1}\sum_r\operatorname{TV}(\widehat P_Q^{A\mid r},\widehat P_{\mathsf H}^{A\mid r})$. For $Q_{16}$, this is $0.004517$, matching the point estimate for independently sampled $\mathsf H'$ against $\mathsf H$; the equality of these estimates does not establish equality of population receiver laws. The empirical conditional decoder information is computed from the same route-balanced contingency tables before and after enforcement, with separate paired-bootstrap uncertainty.

The canonicaliser maps each accepted parsed route to exactly \texttt{domain: <value>} followed by \texttt{severity: <value>}, or maps parser failure to the single string \texttt{REJECT}. It accesses only the incoming text and preserves the receiver's parsed action for every possible input; it does not repair an incorrectly parsed route using the true state. Exhaustive checks on the saved test observations find zero receiver-action changes in $7\times8{,}192=57{,}344$ comparisons. For $Q_{16}$, the hash decoder's recovery changes from $100\%$ to $49.95\%$, and its estimated $I(B;J\mid R)$ falls from $1.0000$ to $0.000629$ bits (a $0.9994$-bit reduction). More generally, with $K=17$ canonical outputs, $C=c(V)$, and task-only canonical target $h(R)$, the observed routing-error fraction $\widehat\varepsilon=115/8192$ gives the empirical-distribution bound $\widehat I(B;C\mid R)\leq h_2(\widehat\varepsilon)+\widehat\varepsilon\log_2(16)\approx0.163$ bits and confines every binary decoder's test-population recovery to $[1/2-\widehat\varepsilon,1/2+\widehat\varepsilon]$. The latter are all-decoder statements under the enforced interface, evaluated on the saved test distribution, not population confidence guarantees.

\subsection{Anonymous Artefact and Reproduction}\label{app-exp-artifact}

\noindent\textbf{Anonymous code artefact.} The source code, tests, frozen configuration, and runner scripts are accessible for anonymous review at \url{https://anonymous.4open.science/r/passive-monitoring-EC82/}. The link is a read-only repository mirror; it is not itself a deposited dataset or a model-weight archive. No identifying source-repository address is needed to follow the study. Reviewers should use the files provided by this anonymous artefact and the documented environment setup; generating a new run and analysing an existing results archive are distinct workflows.

\noindent\textbf{Implementation map.} The repository's \path{README.md} documents setup and supported execution modes, while \path{config/study.json} fixes the checkpoint, draw budgets, split sizes, base seeds, and analysis tolerances. The file \path{study/protocol.py} specifies the $64$ synthetic records, exact prompts, hash decoder, parser, canonicalisation, candidate-selection rule, and stream derivation. Collection and validation are implemented by \path{study/pipeline.py} and the generation engines, including \path{study/production_fast.py}. The primary outputs are computed by \path{study/analysis.py}, with detectors in \path{study/detectors.py} and resampling/statistical utilities in \path{study/metrics.py}. Additional theory-aligned information estimates and receiver-interface checks come from \path{study/theory_analysis.py}; the exact finite witness is independently checked by \path{study/exact.py}. The \path{tests/} directory supplies automated checks of protocol, engines, analysis, and launch behaviour.

\noindent\textbf{Full GPU reproduction.} The documented end-to-end run targets Linux and Python $3.12$, a CUDA GPU with $24$\,GB VRAM, approximately $50$\,GB system RAM, and approximately $25$\,GB available disk space. The reported execution used one NVIDIA RTX A5000, PyTorch $2.8.0$, and CUDA $12.8$; production message collection took approximately $11.5$ hours. After obtaining a local copy of the anonymous artefact, run the following from its root:
\begin{lstlisting}[basicstyle=\ttfamily\scriptsize,breaklines=true,columns=fullflexible]
bash run.sh
tail -f logs/run.log
\end{lstlisting}
The launcher configures the runtime, runs preflight and consistency checks, retrieves the pinned model assets, carries out the frozen generation and analyses, and packages outputs and provenance. The run is launched asynchronously, so \path{logs/run.log} reports progress and completion; the repository README documents restart and checkpoint handling. The locally produced \path{runs/paper_study-results.tar.gz} archive includes collected observations and metadata needed for subsequent CPU-only analyses. It is a \emph{generated output}, not a file assumed to be available in the anonymous code mirror. Repeating inference on other hardware can introduce numerical differences; the artefact retains tolerances, source/configuration hashes, and diagnostic checks rather than promising bitwise portability.

\noindent\textbf{CPU-only checks and reanalysis.} Implementation tests and the frozen run plan can be inspected without generating LLM outputs:
\begin{lstlisting}[basicstyle=\ttfamily\scriptsize,breaklines=true,columns=fullflexible]
bash setup.sh cpu
.venv/bin/python -m pytest -q
.venv/bin/python -m study plan
\end{lstlisting}
For analysis-only reproduction, first obtain the \emph{generated} \path{paper_study-results.tar.gz} archive from a completed production run, place it at the repository root, and unpack it; then execute the two analysis entry points:
\begin{lstlisting}[basicstyle=\ttfamily\scriptsize,breaklines=true,columns=fullflexible]
mkdir -p runs
tar -xzf paper_study-results.tar.gz -C runs
.venv/bin/python -m study analyze \
  --run runs/paper_study \
  --output results/analysis
.venv/bin/python -m study theory \
  --run runs/paper_study \
  --analysis results/analysis \
  --output results/theory
\end{lstlisting}
The primary analysis produces per-budget recovery and task-accuracy summaries, detection results, plots, and supporting records. The supplementary analysis produces \path{public_decoder_information_by_arm.csv}, \path{per_route_information_recovery.csv}, \path{fibre_mass_balance.csv}, \path{task_interface_distances.csv}, \path{canonicalization_information.csv}, and \path{posthoc_theory_aligned_results.json}, together with the underlying detailed tables. Accordingly, a reader can inspect the exact code and run the self-tests from the anonymous artefact, reproduce the full study by rerunning GPU inference, or regenerate analyses rapidly if the corresponding production archive is separately available. These routes should not be conflated: the code mirror alone does not enable instant numerical reproduction of the reported $199{,}680$ generated messages.

\end{document}